\documentclass[preprint,showpacs,preprintnumbers,amsmath,amssymb,nofootinbib]{revtex4}
\usepackage{epsf}
\usepackage{graphics}
\usepackage{graphicx}
\usepackage{dcolumn}
\usepackage{bm}
\usepackage{marvosym}

\renewcommand{\d}{\textrm{d}}
\newcommand{\re}{\scriptsize\textrm{re}}
\newcommand{\im}{\scriptsize\textrm{im}}
\newcommand{\E}{\textrm{E}}
\newcommand{\N}{\textrm{N}}

\newcommand{\paramvec}{\mbox{\boldmath$\theta$}}
\newcommand{\rvec}{\mbox{\boldmath$\tau$}}
\newcommand{\paramvecs}{\mbox{\scriptsize\boldmath$\theta$}}

\newcommand{\trueparamvec}{\mbox{\boldmath$\theta_t$}}
\newcommand{\trueparamvecs}{\mbox{\scriptsize\boldmath$\theta_t$}}

\begin{document}
\title{Bayesian comparison of Post-Newtonian approximations of
  gravitational wave chirp signals}

\author{Richard Umst\"atter}
\email{Richard.Umstaetter@jpl.nasa.gov}
\affiliation{Jet Propulsion Laboratory, California Institute of
             Technology, Pasadena, CA 91109}
\altaffiliation [Also at: ]{LIGO Laboratories, California
  Institute of Technology, Pasadena, CA 91125}

\author{Massimo Tinto}
\email{Massimo.Tinto@jpl.nasa.gov}
\affiliation{Jet Propulsion Laboratory, California Institute of
             Technology, Pasadena, CA 91109}
\altaffiliation [Also at: ]{LIGO Laboratories, California
  Institute of Technology, Pasadena, CA 91125}

\date{\today}

\begin{abstract}  
  We estimate the probability of detecting a gravitational wave signal
  from coalescing compact binaries in simulated data from a
  ground-based interferometer detector of gravitational radiation using
  Bayesian model selection. The simulated waveform of the chirp signal 
  is assumed to be a spin-less Post-Newtonian (PN) waveform of a given 
  expansion order, while the searching template is assumed to be either 
  of the same Post-Newtonian family as the simulated signal or one level 
  below its Post-Newtonian expansion order. 
  Within the Bayesian framework, and by applying a reversible jump Markov 
  chain Monte Carlo simulation algorithm,  we compare PN1.5 vs. PN2.0 and PN3.0 vs. PN3.5 wave 
  forms by deriving the detection probabilities, 
  the statistical uncertainties due to noise as a function of the SNR, 
  and the posterior distributions of the parameters.
  Our analysis indicates that
  the detection probabilities are not compromised when simplified
  models are used for the comparison, while the accuracies in the determination of the parameters
  characterizing these signals can be significantly worsened, no matter
  what the considered Post-Newtonian order expansion comparison is.  
    
\end{abstract}


\maketitle

\section{Introduction}
\label{SecI}
Kilometer-size ground-based interferometric detectors of gravitational
radiation have become operational at several laboratories around the
world \cite{LIGO}, \cite{VIRGO}, \cite{GEO}, \cite{TAMA}. From
locations in the United States of America, Italy, Germany, and Japan,
these instruments have started to search, in the kilohertz frequency
band, for gravitational waves emitted by astrophysical sources such as
spinning neutron stars, supernovae, and coalescing binary systems.

Among the various sources of gravitational radiation that these
instruments will attempt to observe, coalescing binary systems
containing neutron stars and/or black holes are expected to be the
first to be detected and studied. These signals have a unique
signature that enables them to be extracted from wide-band data by
digital filtering techniques \cite{Helstrom68}.  This signature is
their accelerating sweep upwards in frequency as the binary orbit
decays because of energy loss due to the emission of gravitational
radiation.  Coalescing binaries have a potential advantage over other
sources in signal-to-noise ratio (SNR) by a factor that depends on the
square-root of the ratio between the corresponding number of cycles in
the wave trains \cite{Thorne87}.

The standard technique used for extracting these ``chirps'' from the
noisy data is called Matched Filtering. In the presence of
colored noise, represented by a random process $n(t)$, the
noise-weighted inner product between the data stream $d(t)$
recorded by the detector and the gravitational waveform template
$h(t)$ is defined as
\begin{eqnarray}
\langle d, h \rangle &:=& 2\, {\rm Re} \int_{f_L}^{f_U}
\frac{\tilde{d}(f)\, \tilde{h}^*(f)} {S(f)} \,  df \ ,
\label{eq:inner}
\end{eqnarray}
where the symbol $ \tilde{} $ over $d$ and $h$ denotes their Fourier
transform, $S (f)$ is the one-sided power spectral density of the
noise $n(t)$, $f_L$, $f_U$ are the limits of the frequency band of
interest, and the $^*$ represents complex conjugation.  From this
definition the expression of the SNR can be written in the following
form \cite{Helstrom68}
\begin{eqnarray}
{\rm SNR}^2 &:=& \frac{\langle d, h \rangle^2}{{\rm Var}\, \langle n, h
  \rangle} \ .
\label{eq:snr}
\end{eqnarray}
By analyzing the statistics of the ${\rm SNR}^2$, it is possible to
make statements about the presence (or absence) of such a
gravitational wave signal in the data.  This operation of course needs
to be performed over the entire bank of templates over which the SNR
statistics is built upon, since a gravitational wave signal is in
principle determined by a (finite) set of continuous parameters.

The effectiveness of the matched filtering procedure relies on the
assumption of exactly knowing the analytic form of the signal
(possibly) present in the data. Recent breakthroughs in numerical
relativity \cite{P05,CLMZ06,BCCKM06} have started to provide a
complete description of the radiation emitted during the inspiral,
merger, and ring-down phases of generic black hole merger scenarios.
Although the ability of obtaining numerically all the templates needed
in a data analysis search (perhaps hundred of thousands of them) might
be practically impossible, in principle we should be able to compare
these numerically derived waveforms against various analytic templates
obtained under different approximating assumptions. Work in this
direction has already started to appear in the literature
\cite{BCP06,BMWCK06,PBBCKWPM07} within the so called ``frequentist
framework'', in which estimates in the reduction in SNRs and
inaccuracies of the determination of the parameters characterizing the
signal, due to the use of approximated waveforms, have been derived.
Depending on the magnitude of these degradations one can decide
whether to use these approximated waveforms as templates in a data
analysis search.

Since it can be argued that contiguous PN approximations should well
characterize the differences between the ``true signal'' present in
the data and the highest-order PN approximation, in this paper we
perform such a comparison within the Bayesian framework. An analogous,
frequentist analysis has recently been performed by Cutler and
Vallisneri \cite{CutVallis:2007} for the case of super massive black
holes binaries observed by LISA (the Laser Interferometer Space
Antenna).  Their approach relied on the use of the Fisher-Information
matrix, which is known to give good results in the case of large
(hundreds to thousands) SNRs. In the case of ground-based
interferometers instead, since the expected SNRs will be probably
smaller than $10$, a parameter estimation error analysis based on the
Fisher-Information matrix would lead to erroneous results
\cite{Vallis:2007}.

In this paper we will estimate the loss in probability of detection
(i.e. loss of {\it evidence} of a signal to be present in the data) as
a function of the SNR in the following two cases: (i) the true signal
present in the data is a spin-less PN3.5 waveform and the search model
is represented by a spin-less PN3.0, and (ii) the true signal is a
spin-less PN2.0 waveform and the model is a spin-less PN1.5 waveform.
We have limited our analysis to these two separate cases in order to
cover the region of the PN approximations that have already, or are in
the process of, being used in the analysis of the data collected by
presently operated ground-based interferometers.

Our approach relies on a Bayesian Markov Chain Monte Carlo (MCMC)
technique, as MCMC methods have successfully been applied to a large
number of problems involving parameter estimations \cite{Gilks:1996}
in experimental data sets. In our analysis the chirp signals (the one
present in the data and the one used as the model) are characterized
by five parameters: the two masses of the system, $m_1$ and $m_2$,
their time to coalescence $t_C$, the coalescence phase $\phi_C$, and
their distance $r$ from Earth.

Bayesian MCMC methods have already been proven to be capable
of estimating the five parameters of a PN2.0 chirp signal embedded
into noisy data of a single interferometer when using a PN2.0
based model \cite{RoeverMeyerChristensen:2006}, 
and in a coherent search in the time domain for nine parameters using a PN2.5 model 
\cite{RoeverMeyerChristensen:2007} and a PN3.5 model \cite{RoeverMeyerGuidiVicereChristensen:2007} for the phase.
However, it has never been shown before how the resulting evidence and probability distributions
of the parameters are affected by the usage of different PN models.
Here we will estimate the evidence of a signal being buried in noise
and, at the same time, derive the probability distribution of its
parameters when the gravitational wave form of the model is a
simplified version of the signal present in the data.  A Bayesian
analysis naturally justifies Occam's Razor
\cite{Jaynes:2003,Loredo:1992} due to the penalization of unreasonably
complex models by the integration over the parameter space. The paper
is organized as follows. In Section \ref{SecII} we provide a brief
summary of the Bayesian framework and its implementation in our
problem.  After deriving the expressions of the likelihood function
and the priors for the parameters searched for, in Sec.~\ref{SecIII} we
describe the Markov-Chain Monte Carlo sampling technique adopted for
calculating the posterior distributions. 
We then specify in Sec.~\ref{SecIV} the different simulations we have conducted
by introducing the wave forms, noise specifications, and parameter sets.
The final results of our simulations are presented in Sec.~\ref{SecV},
displaying the MCMC based posterior distributions for the
parameters and involved models. The estimated posterior probabilities
of the models are presented as a function of SNR with the corresponding 
uncertainties over the noise realizations. 
We find that the difference in detection probability when using a simplified model rather than 
the true one is negligibly small in comparison to these uncertainties.  
The posterior credibility regions of the parameters reveal offsets from the
true parameter values that can be much larger than the statistical uncertainty, for both
the PN1.5/2.0 and PN3.0/3.5 model comparisons. The PN2.0/2.0 and PN3.5/3.5
comparisons on the other hand, always yield credibility regions that cover the true parameter
values. Finally, in Sec.~\ref{SecVI} we provide our comments and concluding remarks. 

\section{The Bayesian formulation}
\label{SecII}

In this section we will derive the Bayesian full probability model for
our problem, which involves the comparison between the two
possibilities of having either a signal and noise or just noise in the
data. This requires the determination of the likelihoods, the prior
distributions for the parameters associated with the models, and the
resulting posterior distributions.

\subsection{Model definition}
\label{SecIIa}
Let us suppose we observe a data stream $d(t) =
s_{t}(\paramvec;t) + n(t)$ containing the instrumental noise $n(t)$
and a chirp signal $s_{t}(\paramvec;t)$ that we will regard as the
``\emph{true}'' signal. Here, $\paramvec$ is the vector representing
all the parameters associated with the signal, and the noise is
assumed to be a stationary Gaussian random process of zero mean. In
the Fourier domain the observed data can equivalently be written as
$\tilde{d}(\paramvec;f)=\tilde{s}_{t}(\paramvec;f)+\tilde{n}(f)$
(where tilde denotes the Fourier transform operation) and we will
refer to this expression as model $\mathcal{M}_{t}$.

In what follows we will assume the ``\emph{true}'' signal
$\tilde{s}_{t} (\paramvec;f)$ to be the gravitational wave emitted by
a coalescing binary system and represented by a spin-less
Post-Newtonian approximation in phase and Newtonian in amplitude for which
$\paramvec=\{m_1,m_2,r,t_C,\phi_C\}^T$. Here $m_1$ and $m_2$ are the
masses of the rotating objects,  $t_C$  is the coalescence time, $r$
the absolute distance to the binary system, and $\phi_C$ the phase
of the signal at coalescence.

We will then describe the detection and estimation of the parameters
of the ``true'' signal by relying on a spin-less lower-order Post-Newtonian waveform,
$\tilde{s}_{s} (\paramvec;f)$. This simpler model will be referred to as model $\mathcal{M}_{s}$.

The derivation of the detection probability implies a comparison
between model $\mathcal{M}_{s}$ and the null-model, which postulates
mere noise $\tilde{n}(f)$ within the data. This model will bereferred to
as model $\mathcal{M}_n$. Note that no parameter enters into this
model.

\subsection{The likelihood function}
\label{SecIIb}

Since we have assumed the distribution of the random process
associated with the noise of the detector to be Gaussian of zero-mean,
it follows that the likelihood function is proportional to the
exponential of the integral of the squared and normalized residuals
between $\tilde{d}({\trueparamvec};f)$ and signal template
$\tilde{s}(\paramvec;f)$ over the frequency band of interest $[f_L,
f_U]$. We define the data $\tilde{d}(f):=\tilde{d}(\trueparamvec;f)$
to be modeled by $\mathcal{M}_t$ with its ``\emph{true}'' parameter
vector ${\trueparamvec}$. The comparison with the simplified model
$\mathcal{M}_s$, results in the following expression for the
likelihood function
\begin{equation}
p(\tilde{d}|\mathcal{M}_s,\paramvec) \propto \exp \left(-
2\int_{f_L}^{f_U}{\frac{|\tilde{d}(f)-\tilde{s}_{s}(f,\paramvec)|^2}{S_n(f)}
\d f} \right) \ ,
\label{simplemodel1}
\end{equation}
where $S_n(f)$ is the  one-sided  power  spectral density of the
noise. 

By substituting
$\tilde{d}(f)=\tilde{s}_{t}(f,{\trueparamvec})+\tilde{n}(f)$ into
Eq.~\ref{simplemodel1}, the likelihood function becomes 
\begin{equation}
p(\tilde{d}|\mathcal{M}_s,\paramvec) \propto 
\exp \left(-2\int_{f_L}^{f_U}{\frac{|\tilde{s}_{t}(f,{\trueparamvec})
+ \tilde{n}(f)-\tilde{s}_{s}(f,\paramvec)|^2}{S_n(f)} \d f} \right) \ .
\label{simplemodel2}
\end{equation}
In analogy to Eq.~\ref{simplemodel2}, under model
$\mathcal{M}_n$ (with no parameters) the likelihood assumes the following form
\begin{equation}
p(\tilde{d}|\mathcal{M}_n) \propto \exp \left( -
  2\int_{f_L}^{f_U}{\frac{|\tilde{s}_{t}(f,{\trueparamvec})+\tilde{n}(f)|^2}{S_n(f)}
  \d f} \right) \ .
\label{emptymodel1}
\end{equation}

For comparison reasons, in the case of using the ``\emph{true}'' model
the likelihood function becomes
\begin{equation}
p(\tilde{d}|\mathcal{M}_t,\paramvec) 
\propto \exp
\left(-2\int_{f_L}^{f_U}{\frac{|\tilde{s}_{t}(f,{\trueparamvec})+\tilde{n}(f)-\tilde{s}_{t}(f,\paramvec)|^2}{S_n(f)}
    \d f} \right) \ ,
\label{truemodel2}
\end{equation}
which will then give information about the impairment in detection
when using model $\mathcal{M}_s$ instead of $\mathcal{M}_t$. Note,
that the investigation of this question requires to do the two
different model comparisons separately, i.e. $\mathcal{M}_n$ vs.
$\mathcal{M}_s$ and $\mathcal{M}_n$ vs. $\mathcal{M}_t$
\cite{Gregory:2005}.

The next step needed for completing our Bayesian full probability
model is the identification of suitable prior distributions for the
five parameters characterizing the chirp signal.  Our derivation will
closely follow that described in
\cite{RoeverMeyerChristensen:2006,RoeverMeyerChristensen:2007}.

\subsection{Prior distributions}
\label{SecIIc}

The derivation of appropriate priors $p(\paramvec)$ bears significant
influence on the evidence of a signal presence within noise, since the
prior identifies the size of the parameter space which the evidence is
based on.

A detailed description of the derivation of the prior distributions,
and in particular for the masses $m_1$, $m_2$, and distance $r$, can
be found in
\cite{RoeverMeyerChristensen:2006,RoeverMeyerChristensen:2007}.  In
short, the masses are assumed to be uniformly distributed over a
specified range, $[m_{min} , m_{max}]$, and the prior
distribution for the distance is chosen to be a cumulative
distribution of having systems out to a distance $x$ smaller than $r$,
$P(x < r)$, proportional to the cube of the distance, $x^3$. In order
to obtain a proper prior distribution that does not diverge once
integrated to infinity, it is down-weighted by including an
exponential decaying. This accounts for the Malmquist effect
\cite{Sandage:2001} and includes the assumption of uniform
distribution for the masses. The resulting distribution function
$p(m_1, m_2, r)$ can be written as follows \cite{RoeverMeyerChristensen:2006}
\begin{eqnarray}
p(m_1,m_2,r) &\propto& I_{[m_{min},m_{max}]}(m_1)~I_{[m_{min},m_{max}]}(m_2)~r^2 \nonumber \\ 
&& \cdot \left( 1+ \exp \left( \frac{\log \mathcal{A}-a}{b}\right) \right)^{-1}
\label{priorm1m2d}
\end{eqnarray}
where 
\begin{equation}
\mathcal{A}=M^{5/6}_c/r=\frac{\sqrt{m_1 m_2}}{(m_1+m_2)^{1/6}r} \ ,
\end{equation}
and $I{_[m_{min},m_{max}]}(m)$ is the so-called ``Indicator
function'', equal to $1$ when $m_{min} \le m \le m_{max}$ and zero
elsewhere. The latter term containing the log-amplitude in the sigmoid
function of Eq.~\ref{priorm1m2d} is the down weighing term mentioned
earlier, and it depends on two constants $a$ and $b$. These are
determined by requiring a smooth transition of a $m_1, m_2$ inspiral
system being detectable with two specified probabilities at two given
distances. In
\cite{RoeverMeyerChristensen:2006,RoeverMeyerChristensen:2007} $a$ and
$b$ are determined by choosing a $(2-2)$$M_\odot$ inspiral system to be
detectable with probabilities $0.1$ and $0.9$ out to distances $95$
Mpc and $90$ Mpc respectively. In our work we will also make such a
choice.

\begin{figure}[!ht] 
 \begin{center}
  \includegraphics[width=7.0cm,angle=270]{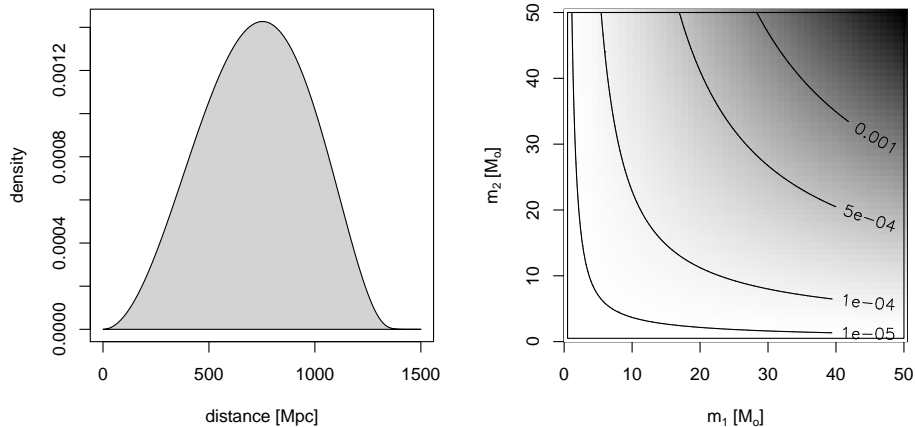}
 \end{center}
\caption{Joint prior distribution of the masses $m_1$, $m_2$, 
  and marginal distribution of the distance $r$, using
  Eq.~\ref{priorm1m2d}. Although initially a uniform distribution is
  assigned to the masses, the conjunction of distance and masses
  results in a higher detectability of large masses. The number of
  possible binary systems increases quadratically with the distance.
  The down weighing of the prior is significantly seen at around
  $500~$Mpc and above as we allow masses of up to $50M_\odot$.}
\label{priorMD}
\end{figure}

Fig.~\ref{priorMD} shows the joint prior distribution of the masses $m_1$, $m_2$, and marginal distribution of the distance $r$, using Eq.~\ref{priorm1m2d}. Although initially a uniform distribution is assigned to the masses, the conjunction of distance and masses 
results in a higher detectability of large masses. The number of possible binary systems increases quadratically with the distance but the down weighing of the prior is significantly seen above $500~$Mpc as we allow masses of up to $50M_\odot$. In \cite{RoeverMeyerChristensen:2006,RoeverMeyerChristensen:2007} for example, the masses where restricted to $3M_\odot$ and therefore the prior values on the distance were much smaller with a distribution mode at around
 $75~$Mpc.  

As far as the time to coalescence is concerned, we have assumed it to
be uniformly distributed over a time interval of $1$ second centered
around the value identified by the masses of the binary and the lower
frequency cut-off of the detector \cite{PDB01}. This search range for
the time to coalescence $t_C$ is larger than that used in
\cite{RoeverMeyerChristensen:2006} because when using the simplified
model the posterior peak can be offset from the true value by more
than the posterior width.

Finally, we chose the phase of the signal at coalescence, 
$\phi_C$, to be uniformly distributed over the interval $[0, 2\pi]$, i.e.
$p(\phi_C)=\textrm{I}_{[0,2\pi[}(\phi_C)/(2\pi)$.

The choice of priors is different when it comes to the analysis of
model $\mathcal{M}_{n}$. Since this model postulates mere noise, there
are no parameters entering the likelihood, which is therefore a
constant.

The final remaining step in defining the Bayesian procedure is to
assign prior probabilities to the models themselves. Since we have no
a priori knowledge, the unbiased choice is equal probability for each.

\subsection{Posterior distribution}
\label{posterior_distribution}

By applying Bayes' theorem using the likelihoods and priors defined
above, we then derive the multidimensional posterior probability
distribution for the model and its parameters
\begin{equation}
p(i,\paramvec_i|\tilde{d})=\frac{\left\{
\begin{array}{ll}
p(\mathcal{M}_n)\cdot p(\tilde{d}|\mathcal{M}_n
) & \textrm{if } i=0\\
p(\mathcal{M}_s,\paramvec)\cdot p(\tilde{d}|\mathcal{M}_s
,\paramvec) & \textrm{if } i=1 
\end{array} \right\}}{\int p(\mathcal{M}_s,\paramvec)\cdot p(\tilde{d}|\mathcal{M}_s,\paramvec
) \d \paramvec + p(\mathcal{M}_n)\cdot p(\tilde{d}|\mathcal{M}_n
)} \ ,
 \label{posterior}
\end{equation}
where $i \in \{0,1\}$ corresponds to the two models
$\{\mathcal{M}_n,\mathcal{M}_s\}$.  In the same way, it is possible to
derive the posterior for the comparison of $\mathcal{M}_n$ vs.
$\mathcal{M}_t$. A Bayesian analysis naturally justifies Occam's Razor
\cite{Jaynes:2003,Loredo:1992} due to the penalization of unreasonably
complex models by integrating over the parameter space resulting in
the preference for a simpler model.

\section{Bayesian model selection and parameter estimation}
\label{SecIII}

There exist various techniques for tackling this multi-dimensional
problem. One possible approach is to calculate the so called Bayes factors 
\cite{Kass:1995,Han:2001,Gregory:2005}, which are the ratios of the
global likelihoods of the models that are involved.
The Bayesian Information Criterion (BIC) \cite{Schwarz:1978} can be used 
as an approximation to the Bayes factor. However, it is possible to address the problem of
sampling from the multidimensional posterior distribution in Eq.~\ref{posterior} 
by implementing a relatively new procedure, called Reversible Jump
MCMC (RJMCMC) technique \cite{Green:1995,GreenHjortRichardson:2003},
which simultaneously addresses the problems of model selection and
parameter estimation. The RJMCMC is combined with traditional fixed
dimension MCMC techniques that sample from the parameters of the
current model. In the following we will briefly review the MCMC
algorithm that we will use in our analysis.

\subsection{Metropolis Coupled Markov chain Monte Carlo}

In Simulated Tempering \cite{Marinari:1992}, the ``temperature''
becomes a dynamic variable on which a random walk is conducted during
the entire sampling process. The joint distribution of temperature and
remaining parameters, however, requires the normalization constants of
the distributions given the temperature. Other approaches like the
Tempered Transition method \cite{Neal:1996} or the Metropolis-Coupled
chain (a.k.a. parallel tempering algorithm) \cite{Geyer:1991} do not
need normalization constants. The latter approach has been advocated
in the astrophysical literature \cite{Gregory:2005} and it has also been
implemented in \cite{RoeverMeyerChristensen:2007,
  UmstaetterThesis:2006}.

In a Metropolis-Coupled chain \cite{Geyer:1991}, sampling is done in
parallel from $k$ different distributions
$p_j(i,\paramvec_i|\tilde{d}), j \in
\{1,\ldots,k\}$. The real posterior distribution of interest is
denoted by $p_j(i,\paramvec_i|\tilde{d})$ with
parameter vector $\paramvec_i$, whereas the distributions of higher
orders $j>1$ are chosen in such way that the sampling process is
facilitated.  Usually, different temperature coefficients are applied
\cite{Hansmann:1997} that flatten out the posterior modes.  During the
sampling from the $k$ distributions, from time to time, attempts are
made to swap the states of a randomly chosen pair of distributions.

The posterior in the present context can be regarded as a canonical distribution
\begin{equation}
p_j(i,\paramvec_i|\tilde{d})=\frac{\left\{
\begin{array}{ll}
p(\mathcal{M}_n)\cdot \exp \left(-
2\beta_j\int_{f_L}^{f_U}{\frac{|\tilde{d}(f)|^2}{S_n(f)}\d
f} \right) & \textrm{if } i=0\\
p(\mathcal{M}_s,\paramvec)\cdot \exp \left(-
2\beta_j\int_{f_L}^{f_U}{\frac{|\tilde{d}(f)-\tilde{s}_{s}(f,\paramvecs)|^2}{S_n(f)}\d
f} \right)
 & \textrm{if } i=1 
\end{array} \right\}}{C}
\label{temperaturescheme}
\end{equation}
where
\begin{eqnarray}
C&=&\int p(\mathcal{M}_s,\paramvec)\cdot \exp \left(-
2\beta_j\int_{f_L}^{f_U}{\frac{|\tilde{d}(f)-\tilde{s}_{s}
(f,\paramvec)|^2}{S_n(f)}\d f} \right) \d \paramvec 
\nonumber 
\\ 
&+& p(\mathcal{M}_n)\cdot p(\mathcal{M}_n)\cdot \exp \left(-
2\beta_j\int_{f_L}^{f_U}{\frac{|\tilde{d}(f)|^2}{S_n(f)}\d
f} \right)
\end{eqnarray}
with inverse temperatures $\beta_j, j \in \{1,\ldots,k\}$. For higher
values of $j$, the posterior modes are flattened out and the sampling
process is eased. A temperature scheme for our Metropolis-Coupled
chain uses $k=10$ different $\beta_j$ values with $j \in \{1,\ldots,k\}$,
where $\beta_1=1$ is the temperature of the original posterior
distribution.  As in
\cite{UmstaetterThesis:2006,RoeverMeyerChristensen:2007}, the prior
distribution is purposely not involved in the temperature scheme as
the prior information at high temperatures is preserved.
Eq.~\ref{temperaturescheme} converges to the prior distribution if
$\beta \rightarrow 0$ whereas a temperature scheme, had it been
applied to the entire posterior distribution, would merely yield a
uniform distribution.

The inverse temperatures $\beta_j, j \in \{1,\ldots,k\}$ are unknown
parameters that must be determined prior to each simulation. It is
obvious that the highest temperature needs to account for the nature
of the likelihood surface. The stronger the signal, the higher the
modes and the more likely it becomes for the MCMC
sampler to get trapped.  The acceptance probability of a proposed jump in a basic
Metropolis-Hastings algorithm is determined by the product of the
ratios between the proposals, the priors, and the likelihoods
\cite{Metropolis:1953,Hastings:1970,Gilks:1996} of the proposed
parameter vector and current state parameter vector. For a coarse
assessment of the nature of the posterior surface, we can neglect the
prior distribution as it is much smoother than the likelihood surface.
With symmetric proposals, the likelihood ratio is therefore the key
factor in analyzing the depth of the modes in the posterior surface.

Although in the simplified model comparison the parameter estimates
can be far off the true parameter values, in the true model comparison
the true parameter values are expected to be good estimates of the
parameters. This fact can be used for a coarse assessment of the
nature of the likelihood surface. Since the log-likelihood of the true
model at the true parameter values is $\log
\rm{LH}_t=-2\int_{f_L}^{f_U}{\frac{|\tilde{d}(f)
    -\tilde{s}_{t}(f,{\trueparamvecs)}|^2}{S_n(f)}\d f}$, it can be
compared to the log-likelihood of the null-model $\log \rm{LH}_n=-
2\int_{f_L}^{f_U}{\frac{|\tilde{d}(f)|^2}{S_n(f)}\d f}$.  The
probability to overcome a proposed MCMC jump between these two
likelihood values determines the convergence of a MCMC sampler. The 
acceptance probability is therefore related to 
the difference of the log-likelihoods $\log
\rm{LH}_t-\log \rm{LH}_n$. We want the hottest temperature to allow
jumps within the posterior surface and we want this to happen about
every, say, $1000$ iterations. This number allows occasional jumps 
at the hottest temperature (smallest
inverse temperature) which is therefore chosen to be
\begin{equation}
\beta_{\min}=\frac{\log(1000)}{\log \rm{LH}_t-\log \rm{LH}_n}.
\end{equation}  
We then use an exponential temperature scheme for $k=10$ chains:
\begin{equation}
\beta_j=\beta_{\min}^{\frac{j-1}{k-1}}, j=\{1,\ldots,k\}.
\end{equation} 

For each iteration and each chain, new parameter values are proposed.
Of course this is only meaningful when the current state of the chain
is not the null-model. If the present state of the sampler is model
$\mathcal{M}_s$ (or $\mathcal{M}_t$ depending on the comparison),
independent normal distributions are chosen to propose new jumps.
Pilot runs are first used to find appropriate proposal variances.  The
acceptance probability for a proposed candidate is derived by
computing the Metropolis-Hastings ratio
\cite{Gilks:1996,Metropolis:1953,Hastings:1970}. The proposed swaps
between arbitrary pairs of chains are done in the way described in
\cite{Geyer:1991} with the temperature scheme highlighted above.

The transdimensional jumps between null-model $\mathcal{M}_n$ and
model $\mathcal{M}_s$ (or $\mathcal{M}_t$) are conducted by RJMCMC
steps \cite{Green:1995}. We implemented a death and a birth proposal
which either attempts to jump from model $\mathcal{M}_s$ (or
$\mathcal{M}_t$) to model $\mathcal{M}_n$ or from model
$\mathcal{M}_n$ to model $\mathcal{M}_s$ (or $\mathcal{M}_t$).

\subsection{Reversible Jump Markov chain Monte Carlo}

The reversible jump approach requires a random variable $\rvec$ with
distribution $q(\rvec)$ that matches the dimensions of the parameter
space across models. In addition, a function is defined that does the
dimension matching. In the present case it is a function based on
death and birth events. The one-to-one transformation in the 'birth'
transition creates a new signal with parameter vector $\paramvec'$ and
has the form $t_{0 \mapsto 1}(\rvec)=\rvec=\paramvec'$. The inverse
'death' transformation that annihilates the signal, $t^{-1}_{0 \mapsto
  1}:=t^{-1}_{1 \mapsto 0}$, has form $t_{1 \mapsto
  0}(\paramvec')=\paramvec'=\rvec$. The Jacobian of both
transformations is equal to $1$. The acceptance probability for the creation
process is therefore
\begin{equation}
\alpha_{0 \mapsto 1}=\min \left\{ 1, \frac{
p(\mathcal{M}_s)p(\paramvec') 
p(\tilde{d}|\mathcal{M}_s,\paramvec') }{
p(\mathcal{M}_n) q(\paramvec') p(\tilde{d}|\mathcal{M}_n)} \right\} ,
\label{accprob01}
\end{equation}
where $\paramvec'=\rvec$ is drawn from $q(\rvec)$. The annihilation process is in turn given by
\begin{equation}
\alpha_{1 \mapsto 0} = \min \left\{ 1,
\frac{p(\mathcal{M}_n)q(\paramvec')p(\tilde{d}|\mathcal{M}_n) }{
p(\mathcal{M}_s)p(\paramvec') p(\tilde{d}|\paramvec',\mathcal{M}_s) } \right\}.
\label{accprob10}
\end{equation}
where $\paramvec'$ is the parameter vector of the current existing
signal.  Since we chose equal prior probabilities for both models, we have
$p(\mathcal{M}_s)/p(\mathcal{M}_n)=p(\mathcal{M}_n)/p(\mathcal{M}_s)=1$.

As one can see in Eq.\ref{accprob01}, for the creation process, the
prior distribution $p(\paramvec')$ at an existing parameter
$\paramvec'$ is found in the enumerator while the proposal
distribution $q(\paramvec')$ at the existing parameter is present in
the denominator.  On the other hand, in Eq.\ref{accprob10}, for the
annihilation process, the proposal value $q(\paramvec)$ of a new
proposed parameter vector $\paramvec'=\rvec$ is found in the
enumerator while the prior $p(\paramvec')$ is in the denominator.
This means that a larger parameter space, which naturally yields
smaller prior values, results in more likely accepted deaths than
accepted births. As a consequence, the sampler will prefer sampling
from the null-model which means that the evidence of a signal will be
smaller if we increase the parameter space. This makes perfect sense
as we expect the evidence of a signal to fade if we integrate over a
larger parameter space.

The other fact that the proposal distribution enters on opposite sides
of the fraction in Eq.\ref{accprob01} and Eq.\ref{accprob10} reveals
the difficulty on the choice of the proposal distribution.  
In order to understand the effect of the proposal distribution,
let us consider the following three scenarios
\begin{enumerate}
\item Suppose we do not know the major posterior mode and therefore
  choose the parameter vector of a new signal to be drawn from a wide
  spread proposal distribution. We could choose the proposal
  distribution to be the same as the prior distribution, in which case
  prior and proposal would cancel out.  From the sampling point of
  view, the samples would account for the prior distribution and the
  acceptance probability would merely contain the likelihood. However,
  it would be unlikely to find the narrow mode in the likelihood
  surface by ineptly poking around in the entire parameter space
  restricted by the prior. Such a RJMCMC sampler would rarely accept
  jumps between the models. Only very long runs would give sufficient
  information about what proportion the sampler naturally stays in
  which model in order to draw reliable conclusions.
\item Suppose we wrongly assume the posterior mode to be concentrated
  in some area of the parameter space far away from the actual
  posterior mode. We would choose the proposal distribution to have
  the major probability in some wrong area of the parameter space.
  Naturally the draws from such distribution would privilege proposals
  in the wrong area of the parameter space but the acceptance
  probability would repress births and support deaths in the area as
  the proposal values would be naturally high. The sampler is balanced
  but mixing would be poor as proposes in the correct area of the
  parameter space are rare, even though their acceptance would be
  facilitated. The mixing would be even worse than in the first
  scenario and even longer runs would be needed to reveal reliable
  information about how long the sampler naturally stays on average in
  which model.
\item Suppose we have a vague idea about where the major posterior mode is
  located and choose our proposal density to be centered around that area.
  The samples would be drawn preferential in that particular area of
  parameter space but the ratio of proposal and prior would
  compensate for that in the acceptance probability. However, the
  likelihood ratio in this area would have a major impact and the
  sampler is more likely to jump between the models revealing the
  proper ratio and model probability in a much shorter sampling
  period.
\end{enumerate}

We see that the choice of a proper proposal distribution is very
important. A good proposal distribution ought to have the major
probability mass concentrated around the expected posterior mode
but with long tails in order to cover the entire prior. 
This is achieved by a mixture distribution between a normal
distribution with small variance and a uniform distribution that covers
the prior range. If we are to compare the
null-model $\mathcal{M}_n$ and the true model $\mathcal{M}_t$, we are
in a lucky position. The mean of the proposal distribution is most
likely to be found around the true parameter values and we only need to find a
variance in the same order of magnitude as the posterior mode which
can be determined by pilot runs.

Things are different when we compare null-model $\mathcal{M}_n$ and
the simplified model $\mathcal{M}_s$. The posterior mode can not be
expected at the true parameter values as the wave form of the
simplified model is definitely not best fit at the true parameter
values. Just augmenting the variance of a proposal distribution 
with mean at the true parameter values would result in bad mixing.
We therefore need pilot runs at higher signal-to-noise ratio in order
to determine the vague center of the posterior mode in the simple
model case. The information acquired from such runs serves to 
determine a suitable proposal distribution. 

\subsection{Within-model Metropolis-Hastings sampling and re-parameterization}

The sampling process of the individual chains when the current state
of the Markov chain is in the model that postulates a signal, is done
by a common MH step \cite{Metropolis:1953,Hastings:1970,Gilks:1996}.
The proposals here is tailored to the expected posterior shape by
choosing a very heavy tailed distribution. This was accomplished by
mixing a normal distribution with exponentially varying variance
\cite{UMDVWC:2004,MaxEnt:2004,UmstaetterThesis:2006}.
 
The high correlation of the mass parameters $m_1$, and $m_2$ in the
posterior distribution needed to be accounted for by re-expressing
them in terms of the following Newtonian and 1.5 PN time to
coalescence \cite{Chronopoulos:2001} 
\begin{equation}
\lambda_1=F_1 \cdot (m_1+m_2)^{-8/3}\frac{(m_1+m_2)^3}{m_1 m_2} \ ,
\end{equation}
\begin{equation}
\lambda_2=F_2 \cdot (m_1+m_2)^{-5/3}\frac{(m_1+m_2)^3}{m_1 m_2} \ ,
\end{equation}
where $F_1=\frac{5}{256}(\pi f_0)^{-8/3}$ and $F_2=\frac{\pi}{8}(\pi
f_0)^{-5/3}$. This turned out to work very well as a
re-parameterization technique for our sampler.

Since $m_1$ and $m_2$ can be written in terms of $\lambda_1$ and
$\lambda_2$ according to the following expressions
\begin{equation}
m_1=\frac{1}{2}\left(C_1-\sqrt{ C_1^2-4 C_2^{1/3}}\right)
\label{m1} \ ,
\end{equation}
\begin{equation}
m_2=\frac{1}{2}\left(C_1+\sqrt{ C_1^2-4 C_2^{1/3}}\right)
\label{m2} \ ,
\end{equation}
where $C_1=\frac{\lambda_2 F_1}{\lambda_1 F_2}$ and
$C_2=\frac{\lambda_2}{F_2}\left(\frac{F_1}{\lambda_1}\right)^4$, it
follows that the Jacobian of this transformation is equal to
\begin{equation}
\det J=-\frac{F_1 F_2 C_2^{1/3} \sqrt{C_1^2-4 C_2^{1/3}}}{(F_1
\lambda_2)^2 - 4(F_2 \lambda_1)^2 C_2^{1/3}}
\end{equation}

Since in the original parameter space we defined a joint density for
\linebreak $\{m_1,m_2,r\}$, the new joint prior distribution of
$\lambda_1$, $\lambda_2$, and $r$ is given by
\begin{equation}
p(\lambda_1,\lambda_2,r)\!\!=\!\!
\left\{
\begin{array}{ll}
p(m_1\!(\!\lambda_1\!,\!\lambda_2\!),\!m_2\!(\!\lambda_1\!,\!\lambda_2\!),\!r)|\det J|
& \textrm{if~~~~} m_{min} \!\le\!m_1(\lambda_1,\lambda_2)\!\le \!m_{max}\\
&  \textrm{and } m_{min}\!\le\!m_2(\lambda_1,\lambda_2)\!\le\! m_{max}\\
0 & \textrm{otherwise}
\end{array} \right.
\end{equation}
where $m_1(\lambda_1,\lambda_2)$ and $m_2(\lambda_1,\lambda_2)$ are
given by Eq.~\ref{m1} and Eq.~\ref{m2}.

The sampling techniques described in the previous subsections are then
used to sample from this new multidimensional parameter space
$\{\emptyset,(\lambda_1,\lambda_2,\phi_C,t_C,r)^T\}$.

\section{Description of the simulations}
\label{SecIV}
For our simulations we have created data sets from ``\emph{true}'' wave
forms of three hypothetical binary inspiral systems (BI).  We will
consider two scenarios where the ``\emph{true}'' wave form is either
of PN 2.0 or PN 3.5 order.  The detection of each scenario is
attempted by either a PN 1.5 or a PN 3.0 wave form respectively.

In the first scenario, PN3.0/3.5, ``\emph{true}'' model $\mathcal{M}_{t}$
containing a PN3.5 wave form is used for creating the observed data 
$\tilde{d}(\paramvec;f)=\tilde{s}_{t}(\paramvec;f)+\tilde{n}(f)$.
In the frequency domain a PN3.5 signal has the form
\cite{Arun:2005,CutVallis:2007}
\begin{equation}
\tilde{s}_{t}(\paramvec;f)=\mathcal{A} f^{-7/6}\exp \left[ i \left(G(\paramvec;f)+H(\paramvec;f) \psi_{3.5}(\paramvec;f) \right)\right]
\label{truesignalA}
\end{equation}
where
\begin{eqnarray}
\psi_{3.5}(\paramvec;f)&=& \alpha_{1.5}(\paramvec;f) + \alpha_{2.0}(\paramvec;f) + \alpha_{2.5}(\paramvec;f) + \alpha_{3.0}(\paramvec;f)  + \alpha_{3.5}(\paramvec;f)
\end{eqnarray}
with
\begin{eqnarray}
\alpha_{1.5}(\paramvec;f)&=&
1+\frac{20}{9} \left( \frac{743}{336}+\frac{11 \mu}{4 M}  \right) \left(\pi M f \right) ^{2/3} -16\pi ^2M f  \nonumber \\
\alpha_{2.0}(\paramvec;f)&=&10 \left( \frac{3058673}{1016064}+\frac{5429 \mu}{1008 M}+\frac{617 \mu^2}{144 M^2}\right) \left( \pi M f \right) ^{4/3} \nonumber \\
\alpha_{2.5}(\paramvec;f)&=&\pi \left( \frac{38645}{756}+\frac{38645}{252}\log \left( \sqrt{6}\left(\pi M f \right) ^{1/3}\right) \right. \nonumber \\
&& \left. - \frac{65\mu}{9M} \left(1+3 \log\left( \sqrt{6}\left(\pi M f \right) ^{1/3}\right)\right)  \right)\left( \pi M f \right) ^{5/3} \nonumber \\
\alpha_{3.0}(\paramvec;f)&=&\left[  \left( \frac{11583231236531}{4694215680}-\frac{640 \pi^2}{3} - \frac{6848 \cdot 0.57721}{21} \right) \right. \nonumber \\
&&  +\frac{\mu}{M} \left(-\frac{15335597827}{3048192} +\frac{2255 \pi^2}{12}  - \frac{1760}{3}\frac{-11831}{9240} +\frac{12320}{9} \frac{-1987}{3080}\right)\nonumber \\
&& \left. +\frac{76055 \mu^2}{1728 M^2} -\frac{127825 \mu^3}{1296 M^3} - \frac{6848}{21}\left( \log(4) \left(\pi M f \right) ^{1/3} \right) \right]\left( \pi M f \right) ^{6/3} \nonumber \\
\alpha_{3.5}(\paramvec;f)&=& \pi \left( \frac{77096675}{254016} + \frac{378515 \mu}{1512 M} - \frac{74045\mu^2}{756 M^2}\right)\left( \pi M f \right) ^{7/3}
\end{eqnarray}
and
\begin{equation}
H(\paramvec;f)=\frac{3}{128}(\pi M_c f)^{-5/3}
\end{equation}
and
\begin{equation}
G(\paramvec;f)=2 \pi f t_C - \phi_C - \pi/4
\end{equation}
with coalescence time $t_C$ and coalescence phase $\phi_C$, involved
masses $m_1$ and $m_2$, total mass $M=m_1 + m_2$, reduced mass
$\mu=m_1 m_2 / M$, and chirp mass $M_c=(m_1^3 m_2^3/M)^{1/5}$. 

The amplitude $\mathcal{A}$ is related to the intensity of
gravitational wave \cite{Chronopoulos:2001} and in the stationary
phase approximation, $\mathcal{A} \propto M_c^{5/6}/r$, where $r$ is
the distance between detector and source. The model is determined by
the five parameters $\paramvec=\{m_1,m_2,r,t_C,\phi_C\}^T$.  The
proportional factor depends on the relative orientation between
detector and source, and we will assume it to be constant and equal to
$1$ since we are more generally interested in the broad assessment of the
evidence of a signal.

The signal used for detection is the 3.0 PN approximation, which will
serve as the simplified model $\mathcal{M}_{s}$. The PN3.0
approximation formulated in the frequency domain is given by the
following expression
\begin{equation}
\tilde{s}_{s}(\paramvec;f)=\mathcal{A} f^{-7/6}\exp \left[ i \left(G(\paramvec;f)+H(\paramvec;f) \psi_{3.0}(\paramvec;f) \right)\right]
\label{simplesignalA}
\end{equation}
where
\begin{eqnarray}
\psi_{3.0}(\paramvec;f)&=& \alpha_{1.5}(\paramvec;f) + \alpha_{2.0}(\paramvec;f) + \alpha_{2.5}(\paramvec;f) + \alpha_{3.0}(\paramvec;f)
\end{eqnarray}

In the exact same way the second scenario involving PN1.5/2.0 wave
forms is approached where the data are created using a supposedly
``\emph{true}'' signal
\begin{equation}
\tilde{s}_{t}(\paramvec;f)=\mathcal{A} f^{-7/6}\exp \left[ i \left(G(\paramvec;f)+H(\paramvec;f) \psi_{2.0}(\paramvec;f) \right)\right]
\end{equation}
where
\begin{eqnarray}
\psi_{2.0}(\paramvec;f)&=& \alpha_{1.5}(\paramvec;f) + \alpha_{2.0}(\paramvec;f).
\end{eqnarray}
The simplified model $\mathcal{M}_{s}$ uses the lower 1.5 PN expansion
\begin{equation}
\tilde{s}_{t}(\paramvec;f)=\mathcal{A} f^{-7/6}\exp \left[ i \left(G(\paramvec;f)+H(\paramvec;f)  \alpha_{1.5}(\paramvec;f) \right)\right]
\end{equation}

The distance of each binary system is varied in order to obtain
different signal-to-noise ratios. The noise realizations are drawn in
such a way that they correspond to the approximated expression 
of the one-sided power spectral density
of initial LIGO cite{Chronopoulos:2001}
\begin{equation}
S_n(f)=\frac{S_0}{5}\left[\left(\frac{f_0}{f}\right)^4+2 \left[ 1+\left(\frac{f}{f_0}\right)^2\right] \right]
\end{equation}
with $S_0=8.0\times 10^{-46}~\textrm{Hz}^{-1}$ being the minimum noise
of the detector and $f_0=175$~Hz the frequency at which the
sensitivity of the detector reaches its maximum.

The noise samples are generated directly from the noise spectrum in
the following way. Let us assume the noise to be white, Gaussian
distributed $\N(0,1)$ (i.i.d.  standard normal at time $t$). Its finite
$[T_s,T_e]$ Fourier transform is given by
\begin{equation}
{\tilde n} (f)=\mathcal{F} \{ n(t)\} = \int_{T_s}^{T_e}{n(t) \exp(-2
  \pi i f t) \d t} \ ,
\end{equation}
which we can write as
\begin{eqnarray}
\mathcal{F} \{n(t)\} &=& 
\int_{T_s}^{T_e}{\!\!\!\!\!\!n(t)\cos(2 \pi f t) \d t}\!-\! i  
\int_{T_s}^{T_e}{\!\!\!\!\!\!n(t) \sin(2 \pi f t) \d t} 
\nonumber \\
&:=& R(f)- i I(f) \ .
\nonumber 
\\
\end{eqnarray}
We aim for deriving expected values and variance of the real part
$R(f)=\int_{T_s}^{T_e}{n(t)\cos(2 \pi f t) \d t}$ and imaginary part
$I(f)=\int_{T_s}^{T_e}{n(t)\sin(2 \pi f t) \d t}$ of the Fourier
transform. Since $n(t)$ is normally distributed with zero mean, it
follows that also the expectation values $\E \left(R(f) \right)=0$ and
$\E \left(I(f) \right)=0$. From this consideration it follows that the
variances of the real and imaginary parts of the Fourier transform of
the noise are equal to
\begin{eqnarray}
\E \left(I^2(f) \right) &=& 
\E \left(\left[ \int_{T_s}^{T_e}{n(t)\sin(2 \pi f t) \d t} \right]
  ^2\right) 
\nonumber 
\\
&=&\E \left( \int_{T_s}^{T_e}
{\int_{T_s}^{T_e}{n(t')n(t'')\sin(2 \pi f t')\sin(2 \pi f t'') \d t' }\d
  t''}\right) 
\nonumber \\
&\stackrel{i.i.d.}{=}&\E \left( \int_{T_s}^{T_e}{n^2(t)\sin^{2}(2 \pi f
    t)\d t}\right) 
\nonumber 
\\
&=& \int_{T_s}^{T_e}{\underbrace{\E(n^2(t))}_{=1}\sin^{2}(2 \pi f t)\d t} \nonumber \\
&=& \int_{T_s}^{T_e}{\sin^{2}(2 \pi f t)\d t}
\end{eqnarray}
and in the same way
\begin{eqnarray}
\E \left(R^2(f) \right) &=& \int_{T_s}^{T_e}{\cos^{2}(2 \pi f t)\d t}
\ ,
\end{eqnarray}
while the expectation value of the product between the real and
imaginary part of the noise is equal to zero.  For large $\Delta
T:=T_e-T_s$ it is $\E \left[I^2(f) \right] \approx \Delta T/2$ and $\E
\left[R^2(f) \right] \approx \Delta T/2$.  Therefore, the samples of
${\tilde n} (f)$ can be generated by sampling ${\tilde n}_{\re}(f)
\sim \N(0,\Delta T/2)$ and ${\tilde n}_{\im}(f)\sim \N(0,\Delta T/2)$
according to a white spectrum of Gaussian noise of variance $1/2$.
These considerations indicate that we can generate the noise samples
directly in the Fourier domain by sampling the real and imaginary
parts of the noise from two independent random number generators that
are Gaussian distributed and have both equal variance $S_h(f)/2$
($S_h(f)$ being the one-sided power spectral density of the noise).

Since the time required to perform a single-signal simulation was
several days on the Jet Propulsion Laboratory Dell XEON cluster
(running 1024 Intel Pentium 4 Xeon processors), we decided to perform
only three simulations for three different binary systems. These were
selected to be close to the ``corners'' of the ($\lambda_1,
\lambda_2$) region of the mass-space, corresponding respectively to
the mass-pairs given in Tab.~\ref{binarysystems}.
\begin{table}
\begin{tabular}{l||c|c|c|c|c|c|c}
system & $M_c$ & $\eta$ & $m_1$ & $m_2$ & $t_C$ & $\phi_C$ & $r$ \\
\hline
B1 &$0.87905M_\odot$&$0.1875$& $~1.8M_\odot$ & $~0.6M_\odot$ & 42.76933s & 0.2 rad & $16-24$Mpc \\
B2 &$3.089506M_\odot$&$0.0112931$& $45.0M_\odot$ & $~0.52M_\odot$ & 5.26426s & 0.2 rad & $22-35$Mpc \\
B3 &$31.85576M_\odot$&$0.24$& $45.0M_\odot$ & $30.0M_\odot$ & 0.10778s & 0.2 rad & $100-250$Mpc \\
\end{tabular}
\caption{Table of the parameters of the three example binary systems B1-B3.}
\label{binarysystems}
\end{table}
The reason why we chose binary systems of such particular mass constellations can be seen in the following
Fig.~\ref{triangle} where the masses are drawn in the re-parameterized $(\lambda_1,\lambda_2)$-plane. 
\begin{figure}[!ht] 
 \begin{center}
  \includegraphics[width=7.0cm,angle=270]{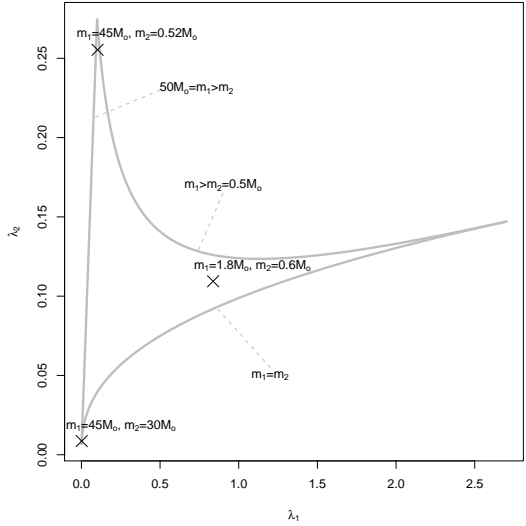}
 \end{center}
\caption{Three example binary systems in the $\lambda_1,\lambda_2$ plane.}
\label{triangle}
\end{figure}
Here, the chirp mass is defined as $M_c=(m_1^3 m_2^3/M)^{1/5}$ and the
mass function $\eta=\frac{m_1 m_2}{(m_1+m_2)^2}$, which is the ratio
between the reduced mass and the total mass of the binary system.
Fig.~\ref{triangle} shows the three mass constellations close to the
three corners of the $(\lambda_1,\lambda_2)$ triangular plane.  The
bottom left corner corresponds to large masses with low mass ratio, in
the top left corner mass constellations are found with large and small
masses (large mass ratio). Towards the right corner, the masses become
small.

\section{Results}
\label{SecV}
We created data sets for the three different binary systems given in
Tab. \ref{binarysystems}, and changed their distances in such a way
that the resulting SNRs would give a detection probability varying
within its extremes. This resulted in varying the SNR within the
interval ($3, 12$).  For the binary system B1, we simulated 12
different distances of varying step width between $16$ and $24~$Mpc.
We found this step width to be sufficient in order to capture the
variability of the calculated detection probability as a function of
the SNR. For binary system B2 we similarly took 12 different distances
between $22$ and $35~$Mpc, while for system B3 we considered 16
different distances in the range of $100-250~$Mpc in steps of
$10~$Mpc. For each of the 40 distances considered we generated 20
different noise realizations, resulting in a total of 800 data sets.
Our MCMC sampler was applied four times on each data set for covering
the specified model comparisons. This yields a total of $3200\times
10$ simulated Metropolis-Coupled MCMC chains, each of which was
stopped after $300\,000$ iterations after inspecting that such a
number was sufficient for our purpose.

The simulated data were sampled at $4096~$Hz for a duration of about
$24~$s, and they were produced by embedding the different signals into
noise samples that where generated in the Fourier domain as described
in Sec.\ref{SecIV}. Since all MCMC runs were conducted after pilot runs at higher SNRs,
the burn-in period was kept very short as the proposal distributions
were optimized to the target distribution and mixing was very
efficient.  From the MCMC output we discarded just the first
$10\,000$ iterations as burn-in, while short-term correlations in the
chain were eliminated by ``thinning'' the remaining terms: every
100$^{\textrm{th}}$ item was kept in the chain.

The integration bandwidth for the likelihood was chosen from $12~$Hz
up to the frequency of the last stable orbit or $600~$Hz, whichever is
the smaller.  Since the SNR is negligible above $600~$Hz, we fixed
this to be the upper frequency cut-off.  For the B1 system, we derived
a frequency of $1832.2~$Hz at the last stable orbit which gives an
integration limit of $600~$Hz for B1. This translates in $14355$
complex samples that contribute to the posterior distribution.  In the
case of system B2 instead, the last stable orbit is at $96.6~$Hz,
resulting in $2065$ complex samples involved in the determination of
the likelihood function.  Finally, for the high mass binary system B3
in our set of systems the last stable orbit is at $58.6~$Hz, implying
now only $1138$ complex samples over which the likelihood is
calculated.

\begin{figure}[!ht] 
 \begin{center}
\begin{tabular}{c}
    \includegraphics[width=8.5cm,angle=270]{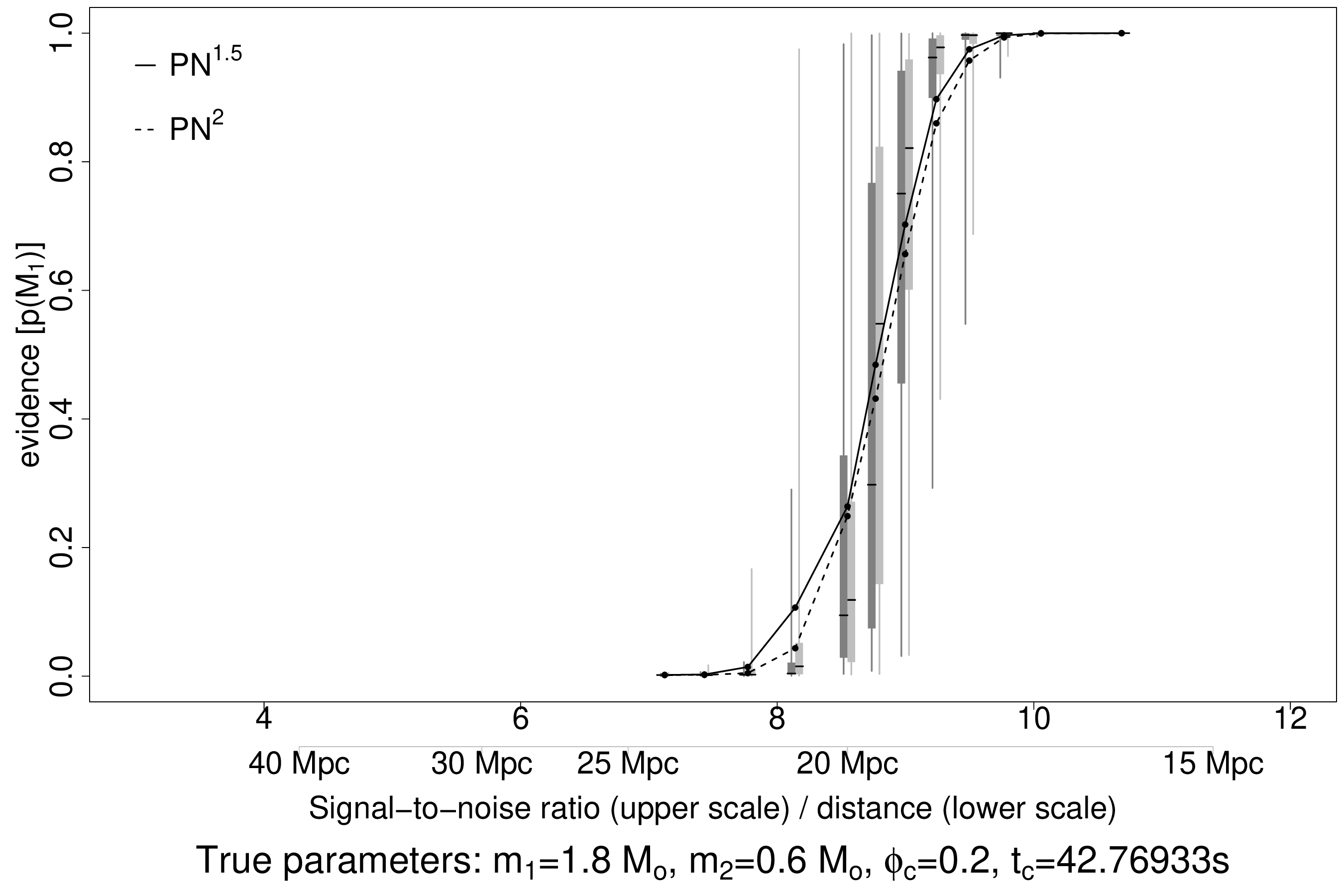}\\
    (a)\\
    \includegraphics[width=8.5cm,angle=270]{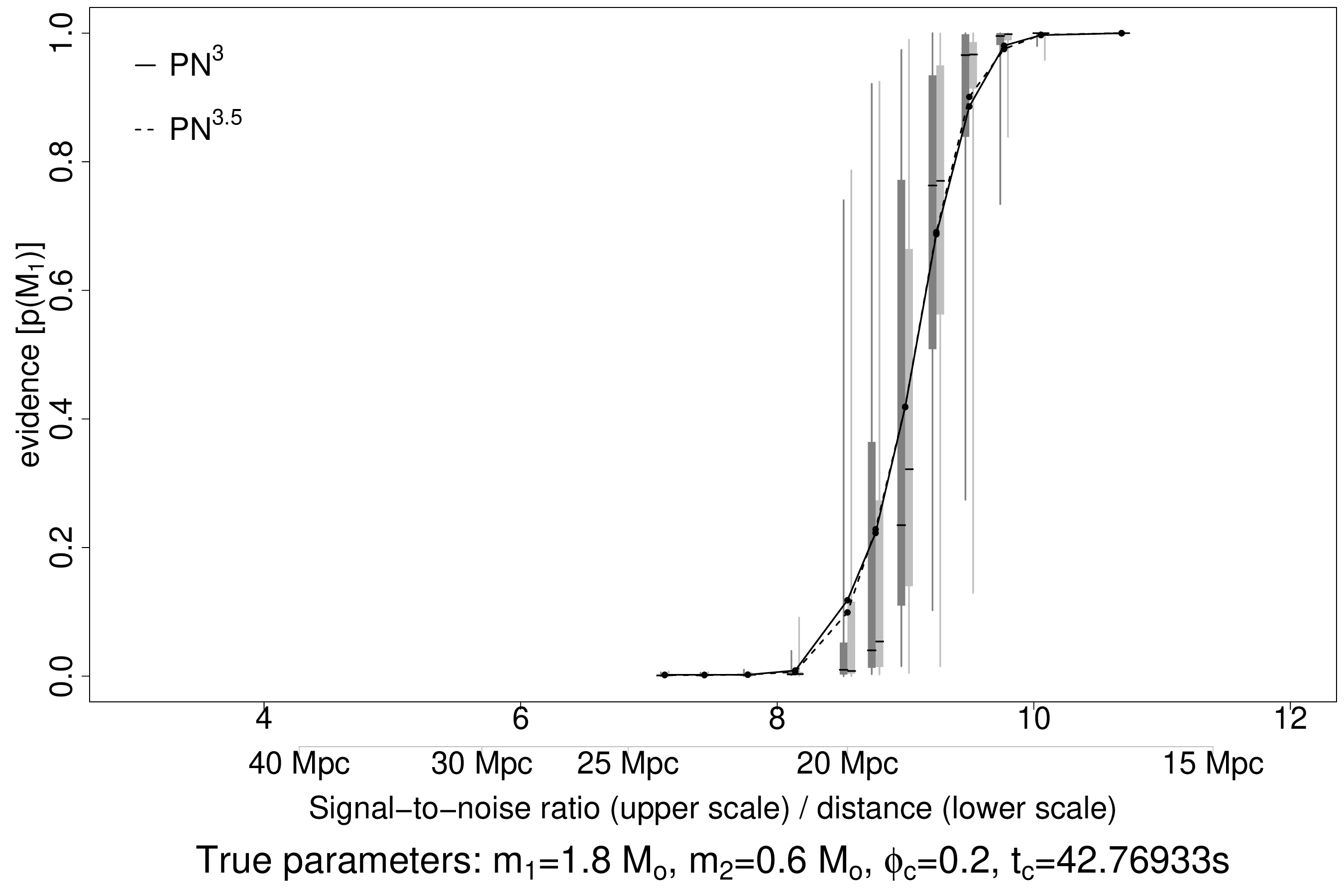}\\
    (b)
        \end{tabular}
  \end{center}
\caption{Probability detection curves of B1 for the PN1.5/2.0 and PN2.0/2.0 comparisons (a) and the 
PN3.0/3.5 and PN3.5/3.5 comparisons (b). 
The vertical gray bars indicate the 50\% quartiles, and the thin lines refer to the outer quartiles
associated with the 20 noise realizations. The inner 50\% quartiles
are divided by a small line which represents the median. 
The lower PN vs. higher PN comparisons are shown as solid lines (detection curves) and light gray bars (quartiles) while the  equal PN comparisons are displayed as dashed lines and dark gray quartile bars. }
\label{EvidenceB1}
\end{figure}

\begin{figure}[!ht] 
 \begin{center}
\begin{tabular}{c}
    \includegraphics[width=8.5cm,angle=270]{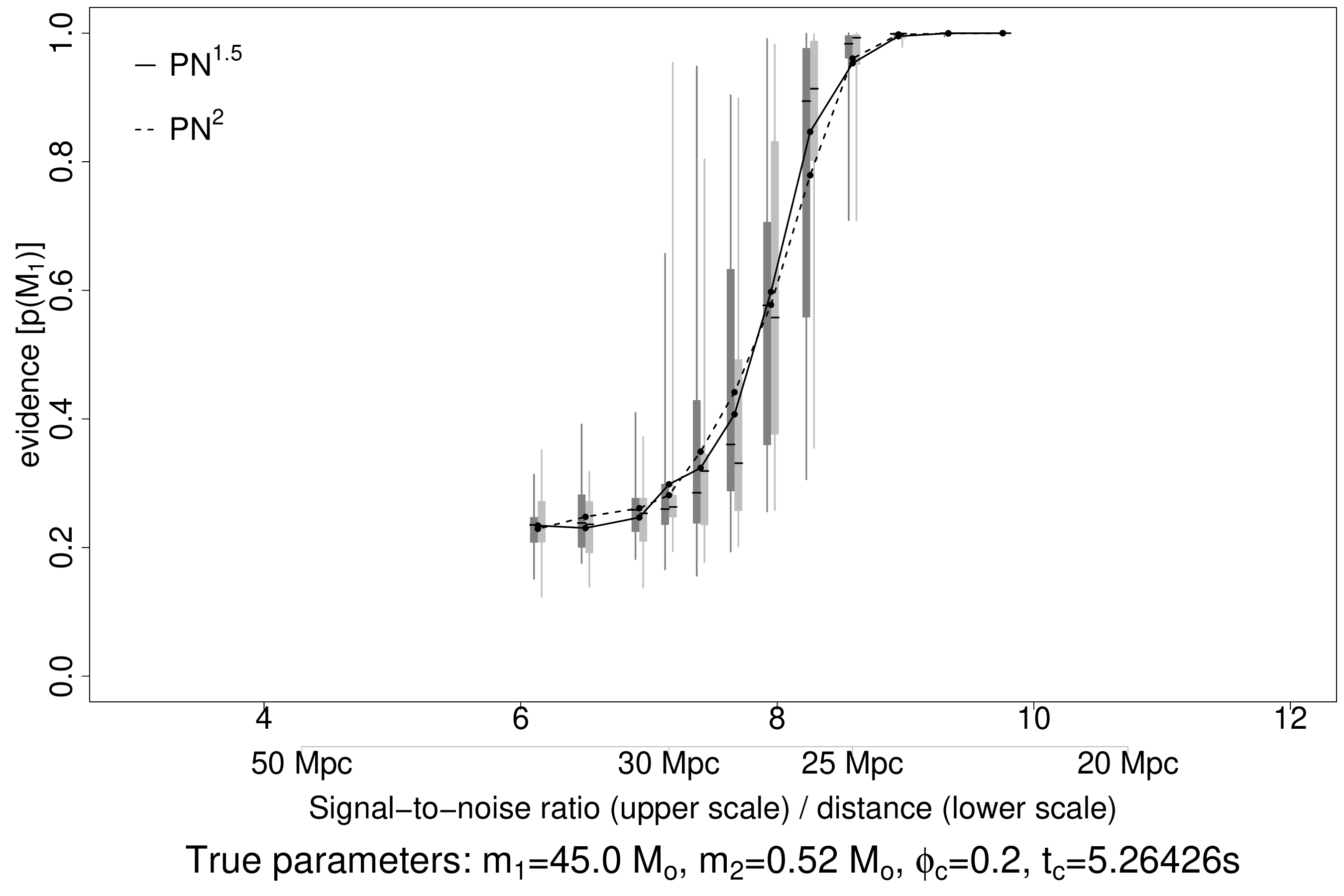}\\
    (a)\\
     \includegraphics[width=8.5cm,angle=270]{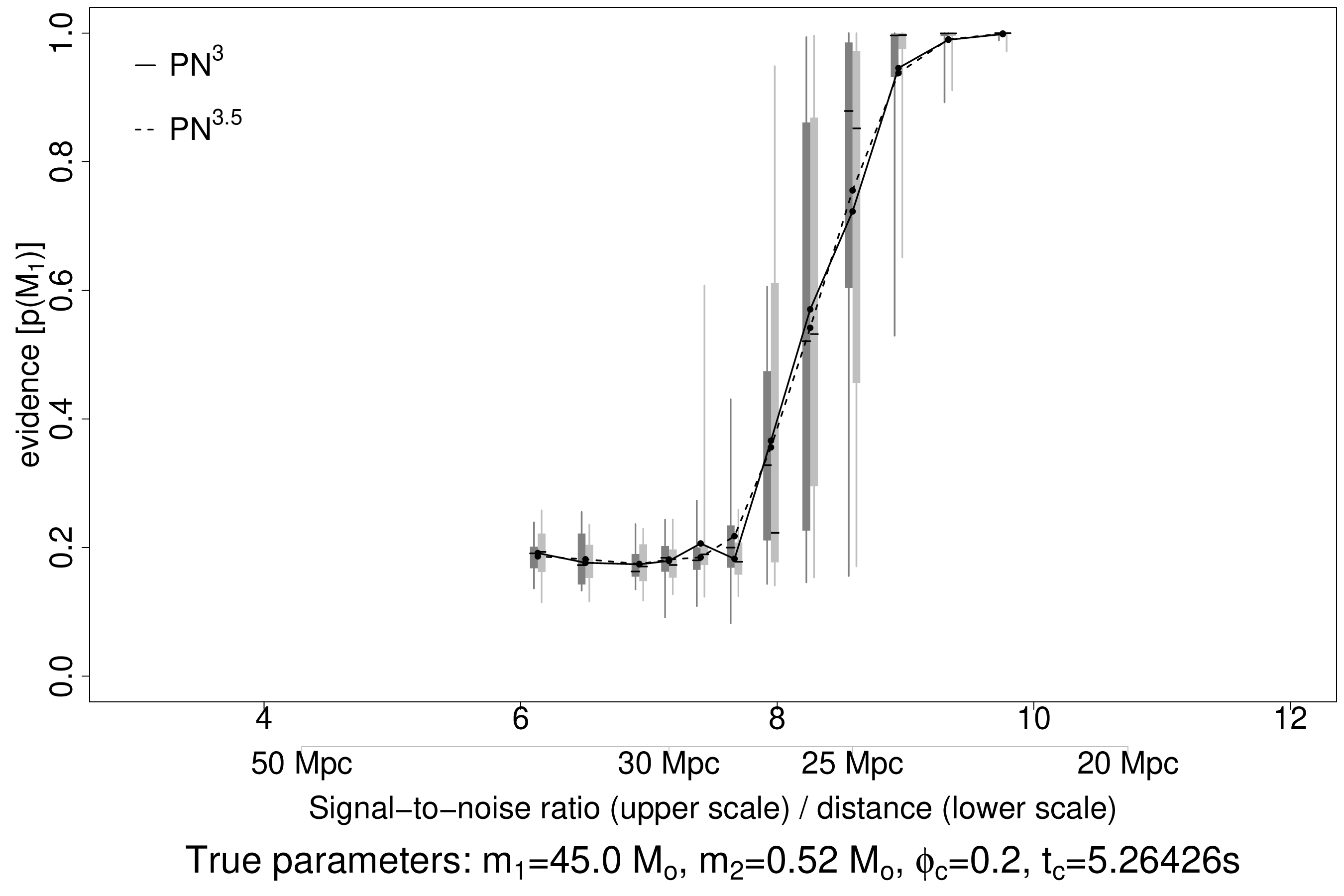}\\
     (b)
     \end{tabular}
  \end{center}
\caption{Probability detection curves of B2 for the PN1.5/2.0 and PN2.0/2.0 comparisons (a) and the 
PN3.0/3.5 and PN3.5/3.5 comparisons (b). 
The vertical gray bars indicate the 50\% quartiles, and the thin lines refer to the outer quartiles
associated with the 20 noise realizations. The inner 50\% quartiles
are divided by a small line which represents the median. 
The lower PN vs. higher PN comparisons are shown as solid lines (detection curves) and light gray bars (quartiles) while the  equal PN comparisons are displayed as dashed lines and dark gray quartile bars. }
\label{EvidenceB2}
\end{figure}

\begin{figure}[!ht] 
 \begin{center}
\begin{tabular}{c}
    \includegraphics[width=8.5cm,angle=270]{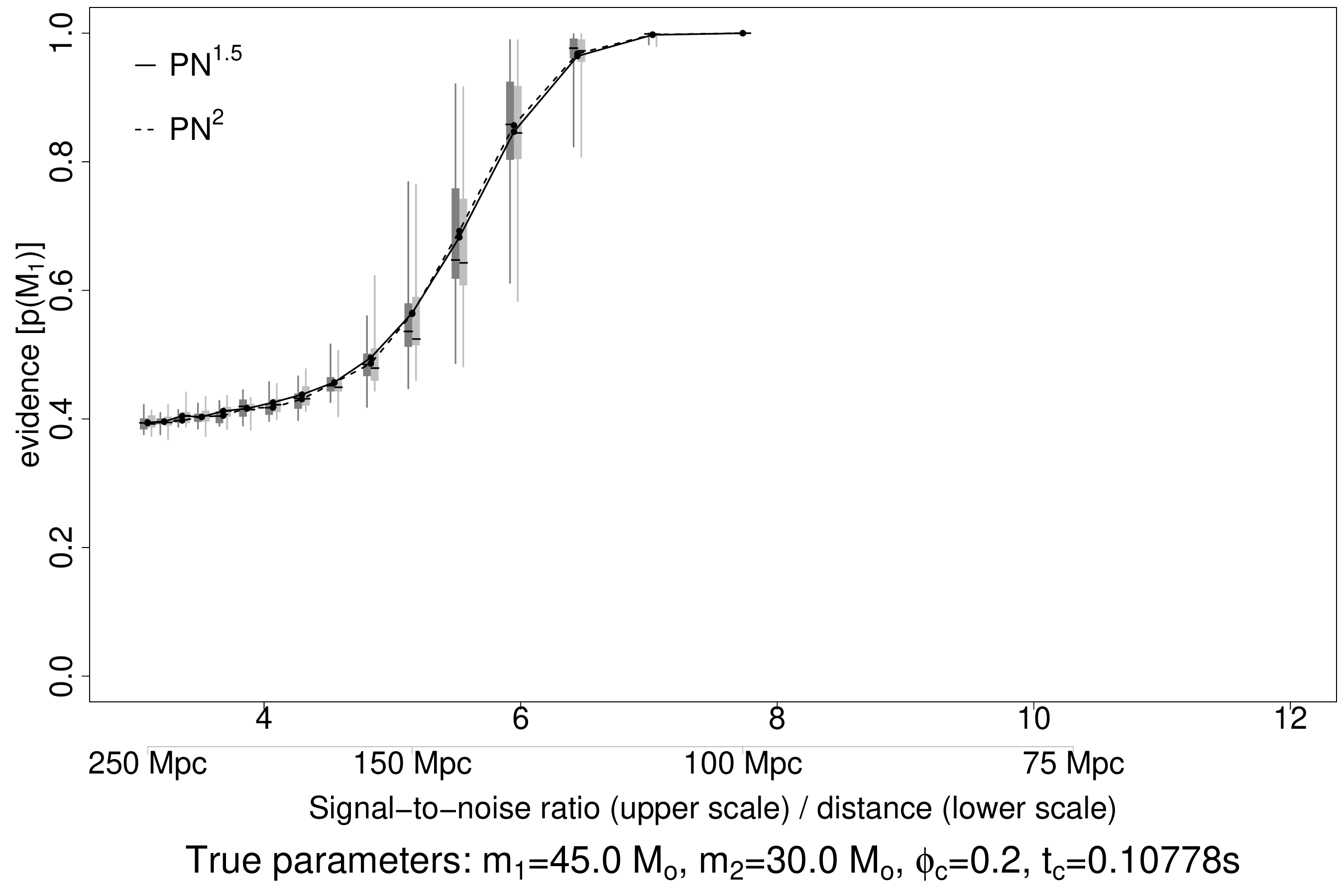}\\
    (a)\\
    \includegraphics[width=8.5cm,angle=270]{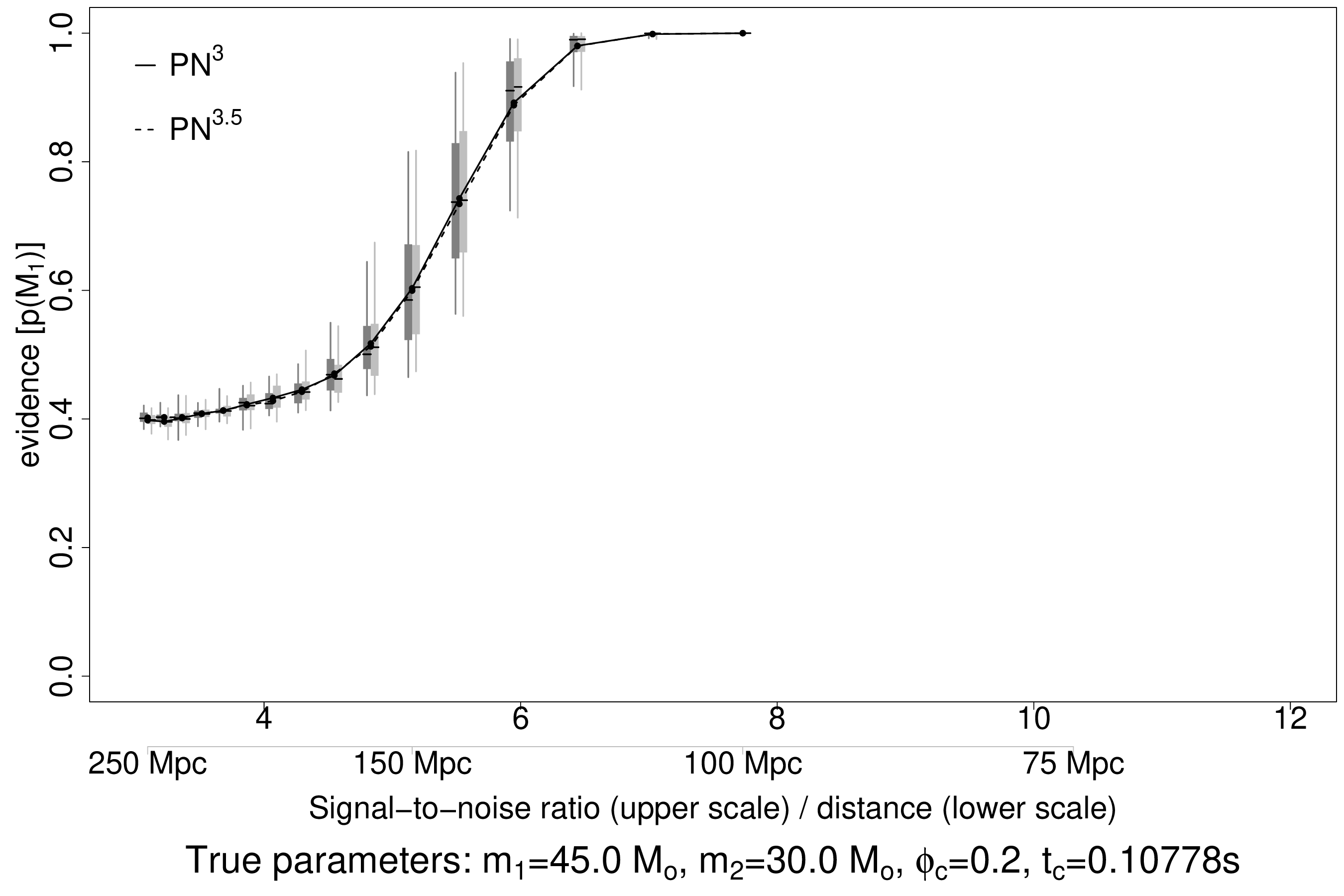} \\
    (b)
        \end{tabular}
  \end{center}
\caption{Probability detection curves of B3 for the PN1.5/2.0 and PN2.0/2.0 comparisons (a) and the 
PN3.0/3.5 and PN3.5/3.5 comparisons (b). 
The vertical gray bars indicate the 50\% quartiles, and the thin lines refer to the outer quartiles
associated with the 20 noise realizations. The inner 50\% quartiles
are divided by a small line which represents the median. 
The lower PN vs. higher PN comparisons are shown as solid lines (detection curves) and light gray bars (quartiles) while the  equal PN comparisons are displayed as dashed lines and dark gray quartile bars. }
\label{EvidenceB3}
\end{figure}

After we conducted the MCMC simulations we derived the posterior
detection probabilities for the competing models from the MCMC outputs
regarding the three example binary systems, the four different model
comparisons, and the different sets of SNRs. For each binary system we
computed the posterior probabilities for the considered scenarios and
contrasted the probability of detection based on a lower order PN
expansion against the one using the true wave form.  This is
displayed in Figs.~\ref{EvidenceB1},~\ref{EvidenceB2},~\ref{EvidenceB3} respectively.

In order to derive the detection curves we computed the proportions of
the states in which the Markov chain recurred to the null model or the
model containing a signal. This was done for the entire set of noise
realizations for a given SNR. Lines connect the estimates of the
posterior detection probabilities resulting in an interpolated
function of the SNR.  This is a monotonically increasing function of
the SNR, reaching asymptotically $1$ as the SNR goes to infinity.  A
common feature to these figures is the uncertainties due to the noise
that the signal detection probability shows at a given SNR. In
particular, these uncertainties are more pronounced when the gradient
of the detection probabilities is at its maximum. Note also that the
difference between the detection probabilities associated to the
``true model'' and the approximated one is much smaller than these
uncertainties.  The uncertainties are displayed as vertical bars: the
50\% quartiles (thick bars), and the outer quartiles (thin lines)
associated with the 20 noise realizations. The inner 50\% quartiles
are divided by a small line which represents the median.  It is
interesting to see in Fig.~\ref{EvidenceB1} that, in some cases, for a
given binary system and SNR, detection probabilities as low as $0$ or
as high as $1$ are possible, merely on the effect of the noise
realization.

It is worth mentioning that analyses performed within the frequentist
framework \cite{Chronopoulos:2001,Canitrot:2001,PBBCKWPM07} and aimed at comparing
the detectability of a signal by using a simplified wave form were
focused entirely on estimating the resulting loss of SNR. The Bayesian
model comparison presented here has the inherent ability to estimate
probabilities and their uncertainties due to noise, providing much
more insights into this issue.

Another interesting feature shown by the detection probability curves
is their asymptotic dependence on the SNR. While the probability of
detection always converges to $1$ as the SNR goes to infinity, it does
not necessarily always goes to zero with the SNR. The reason for this
lies in the Bayesian approach in which we assumed equal prior
probability for both, the null-model and the model that contains a
signal. In the Bayesian context, all probabilities represent a degree
of belief. They are based on the prior information and on the
information that is given by the data by means of the likelihood. The
more data we have, the more new information we obtain from the
posterior distribution. The less data we have, the more impact the
prior has on the posterior. In an extreme scenario with no data at
all, the posterior is equal to the prior. We used this fact to test
the correctness of our RJMCMC sampler as it must sample properly from
the prior distributions and prior model probabilities when no data are
present.  The three binary systems considered in this paper each imply
different likelihoods. For example B1, with its small masses, has a
spectrum that nicely falls into the part of the observable band of the
detector where the instrumental noise is at its minimum. On the other
hand, the system B3 shows an energy spectrum whose upper frequency
cut-off is equal to $58.6~$Hz, with a resulting $1138$ frequency bins
over which the likelihood is calculated. 

The diverse data sets are reflected in
Figs.~\ref{EvidenceB1},~\ref{EvidenceB2},~\ref{EvidenceB3}.  For B1
with its $14355$ involved samples, the detection probability converges
to almost zero for low SNRs as the data provide sufficient evidence
for the non-existence of a signal even though the prior suggests a
probability of $0.5$. It is in the nature of the Bayesian approach
that the scarcer data for B2 and B3 provide less evidence resulting in
a posterior detection probability of around $0.2$ and $0.4$,
respectively, when the SNR approaches zero. A different choice for the
prior probability on $P(\mathcal{M}_n)=1-P(\mathcal{M}_s)$ would
change the course of the posterior detection curves but with
increasing number of data samples and SNR, the likelihood dominates
the posterior distribution.

\begin{figure}[!ht] 
 \begin{center}
    \includegraphics[width=10cm,angle=270]{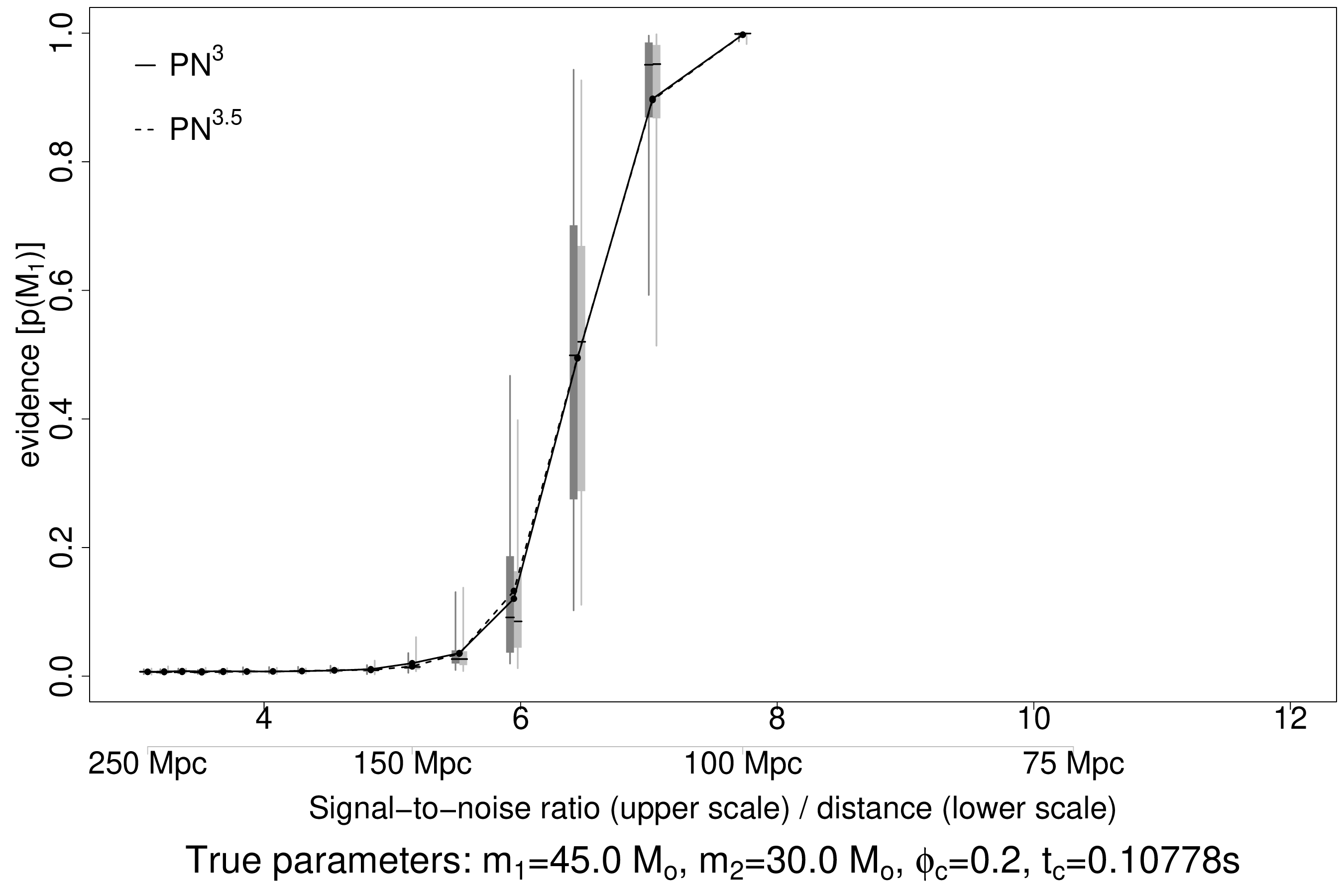}\\
  \end{center}
\caption{Probability detection curves of B3 for the
PN3.0/3.5 and PN3.5/3.5 comparisons. 
The vertical gray bars indicate the 50\% quartiles, and the thin lines refer to the outer quartiles
associated with the 20 noise realizations. The inner 50\% quartiles
are divided by a small line which represents the median. 
The lower PN vs. higher PN comparisons are shown as solid lines (detection curves) and light gray bars (quartiles) while the  equal PN comparisons are displayed as dashed lines and dark gray quartile bars. }
\label{EvidencePrior}
\end{figure}

In order to point this up, we created a graph comparable to
Fig.~\ref{EvidenceB3} (bottom) in which the results of B3 based on the
PN3.0/3.5 comparison are shown with a different prior probability of
$P(\mathcal{M}_n)=0.99$ for the null model. The model that contains a
signal has consequently a prior probability of $0.01$.The result is
illustrated in Fig.~\ref{EvidencePrior}.
This plot corresponds to bottom graph of Fig.~\ref{EvidenceB3} with the only 
difference that a more pessimistic prior probability $P(\mathcal{M}_n)=0.99$ on the null model
has been applied.
When comparing Fig.~\ref{EvidencePrior} to Fig.~\ref{EvidenceB3}
(bottom), we see that the detection probability is significantly lower
at SNR$~<~7$ due to the higher prior probability on the null model.
However, at an SNR of around $6$, the detection curve jumps up quickly
in Fig.~\ref{EvidencePrior} and a detection probability of $1$ is
reached in both figures roughly at an SNR of $8$ because the evidence
of a signal in the data is overruling the prior probability in both
cases.

We will now focus on the parameter estimates. We have compiled plots
which address the impact of the use of lower PN order wave forms on
the bias of posterior distributions of the parameters.  Along the
lines of Figs.~\ref{EvidenceB1},~\ref{EvidenceB2},~\ref{EvidenceB3}, we
plot the posterior distributions of the parameters when the posterior
detection probability reaches a value of $1$.  The posterior
distribution of the parameters is hereby an integration over the noise
by incorporating all $20$ realizations of the MCMC outputs.  Each
output corresponds to the SNRs at which the detection probabilities in
Figs.~\ref{EvidenceB1},~\ref{EvidenceB2},~\ref{EvidenceB3} reach their
maximum. We only concentrated on the four parameters $m_1$, $m_2$,
$r$, $t_C$ as the phase $\phi_C$ is of no particular physical
interest.  We displayed the posterior density of the masses as a 2D
joint probability density in the form of a contour plot. We computed the two-dimensional
50\% and 95\% credibility regions. 

\begin{figure}[!ht]
 \begin{center}
 \begin{tabular}{c}
    \includegraphics[width=7.35cm,angle=270]{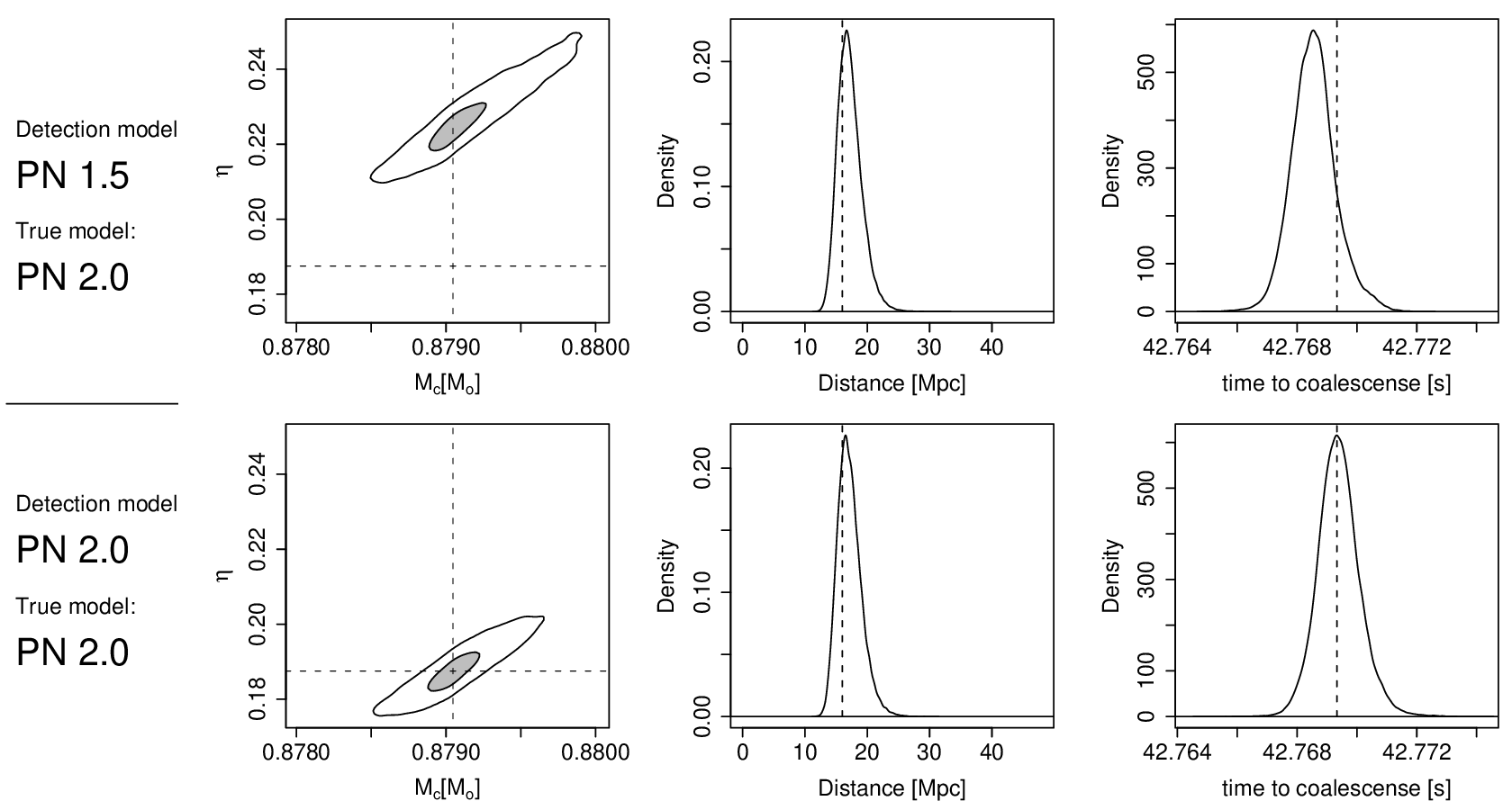}\\
    (a)\\
    \includegraphics[width=7.35cm,angle=270]{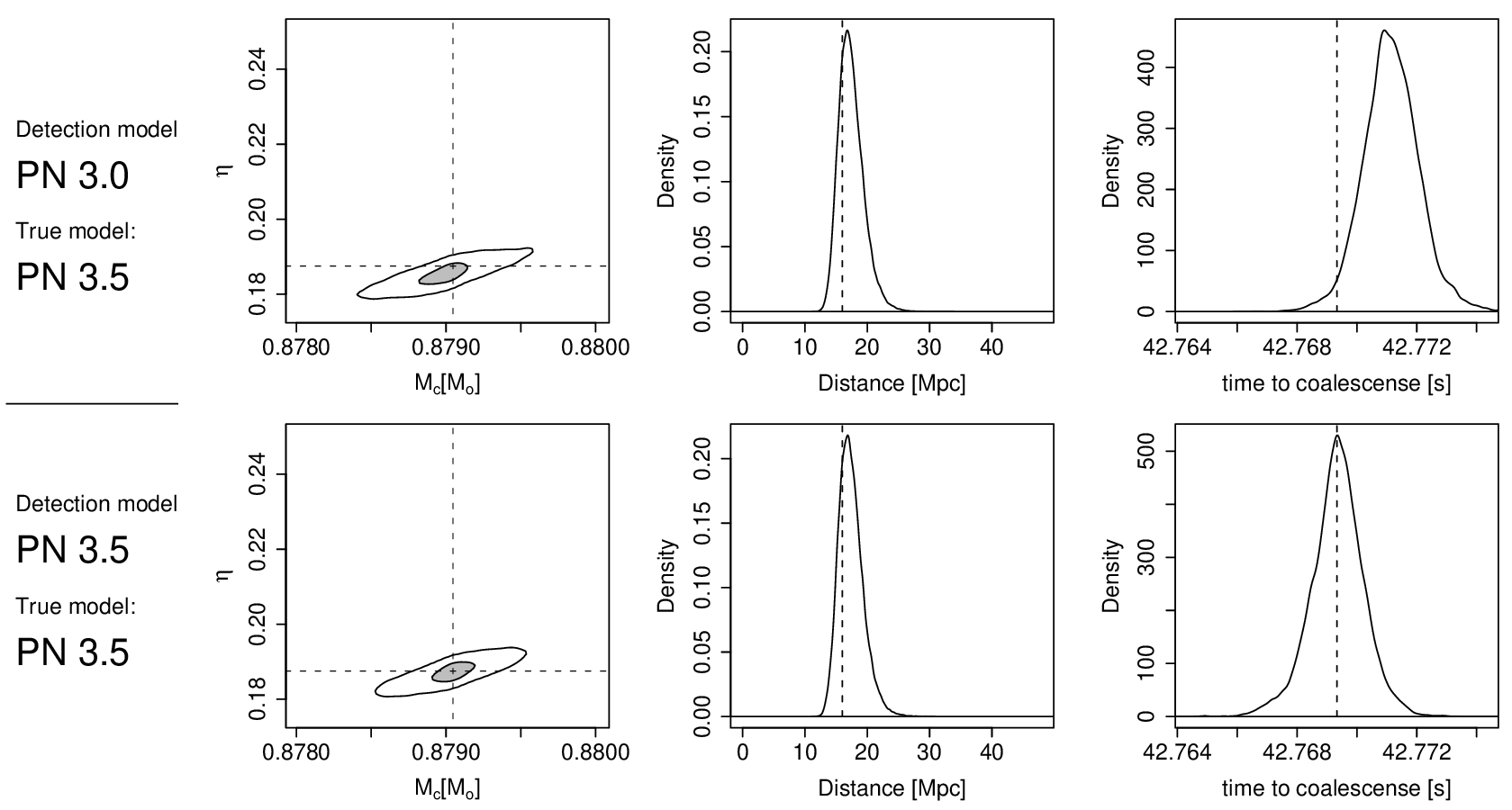}\\
    (b)
 \end{tabular}
\caption{MCMC generated posterior densities for B1. Part (a) shows the comparison with data of a PN2.0 wave form and (b) uses data with the PN3.5 wave form. 
Each of the figures in (a) and (b) show two different model comparisons based either on a lower PN order signal or the same PN order that was used in the data.
The left column shows the joint posterior density of the mass parameters $M_c$ and $\eta$
in form of the 95\% credibility area that contains 95\% of the probability mass and the inner 50\% 
credibility region colored in gray. The middle column shows the MCMC generated kernel density estimate (KDE) of the distance $r$ and the right column the KDE for the time to coalescence $t_C$. The true parameters values are indicated as dashed lines.}
\label{figB1}
\end{center}
\end{figure}

\begin{figure}[!ht]
 \begin{center}
  \begin{tabular}{c}
    \includegraphics[width=7.35cm,angle=270]{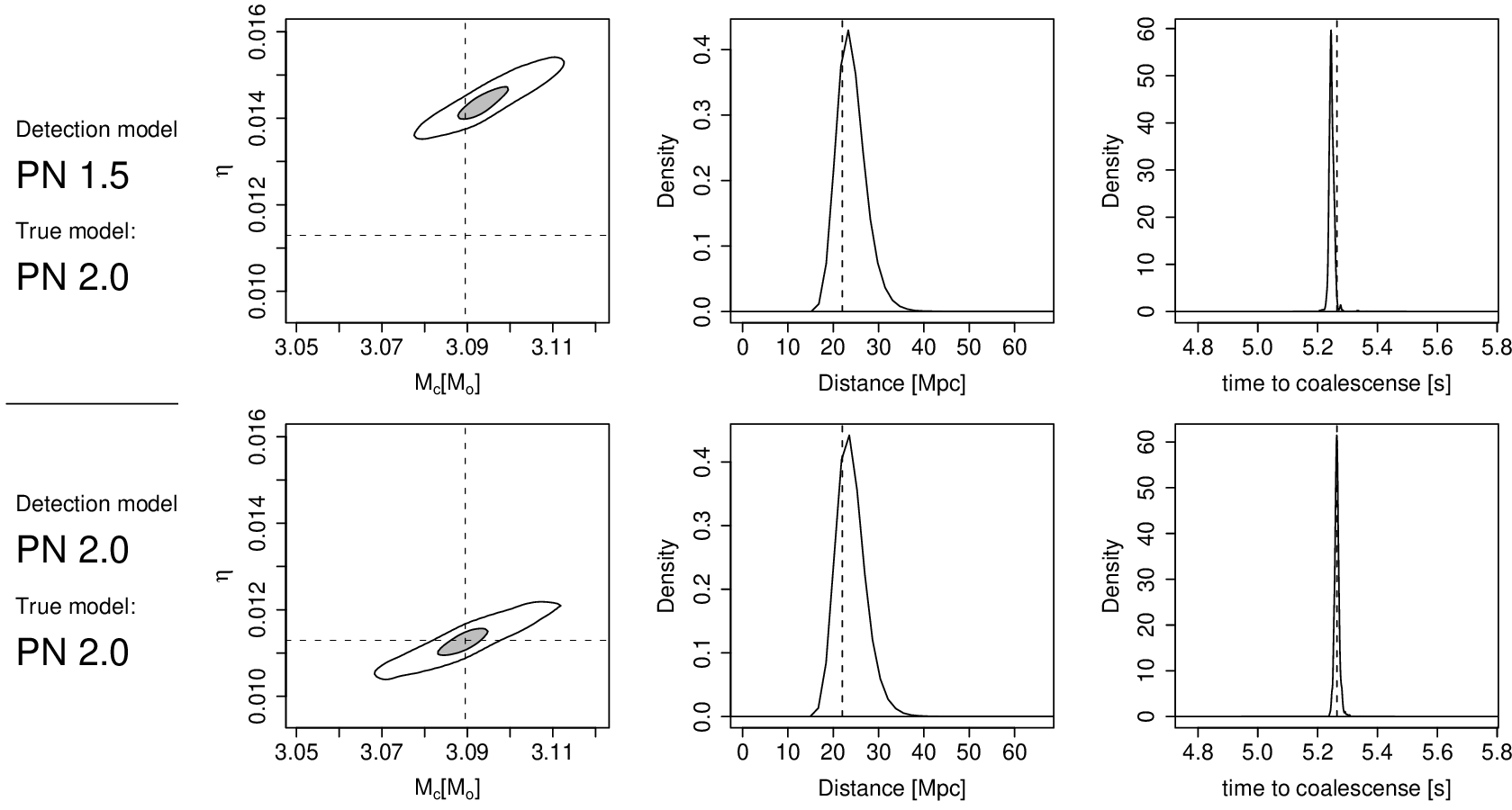}\\
    (a)\\
    \includegraphics[width=7.35cm,angle=270]{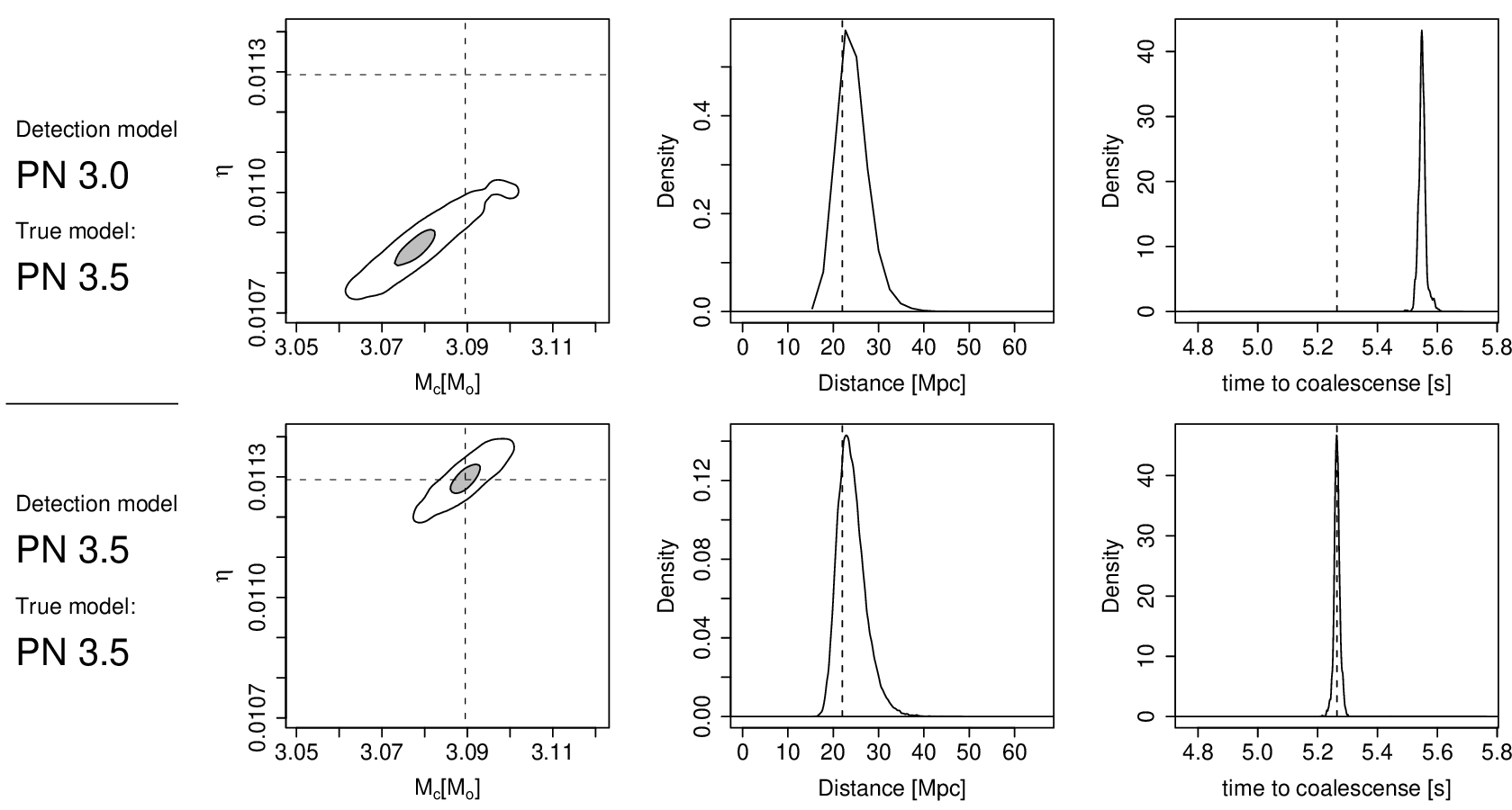}\\
    (b)
   \end{tabular}
\caption{MCMC generated posterior densities for B2. Part (a) shows the comparison with data of a PN2.0 wave form and (b) uses data with the PN3.5 wave form. 
Each of the figures in (a) and (b) show two different model comparisons based either on a lower PN order signal or the same PN order that was used in the data.
The left column shows the joint posterior density of the mass parameters $M_c$ and $\eta$
in form of the 95\% credibility area that contains 95\% of the probability mass and the inner 50\% 
credibility region colored in gray. The middle column shows the MCMC generated kernel density estimate (KDE) of the distance $r$ and the right column the KDE for the time to coalescence $t_C$. The true parameters values are indicated as dashed lines.}
\label{figB2}
\end{center}
\end{figure}

\begin{figure}[!ht]
 \begin{center}
  \begin{tabular}{c}
   \includegraphics[width=7.35cm,angle=270]{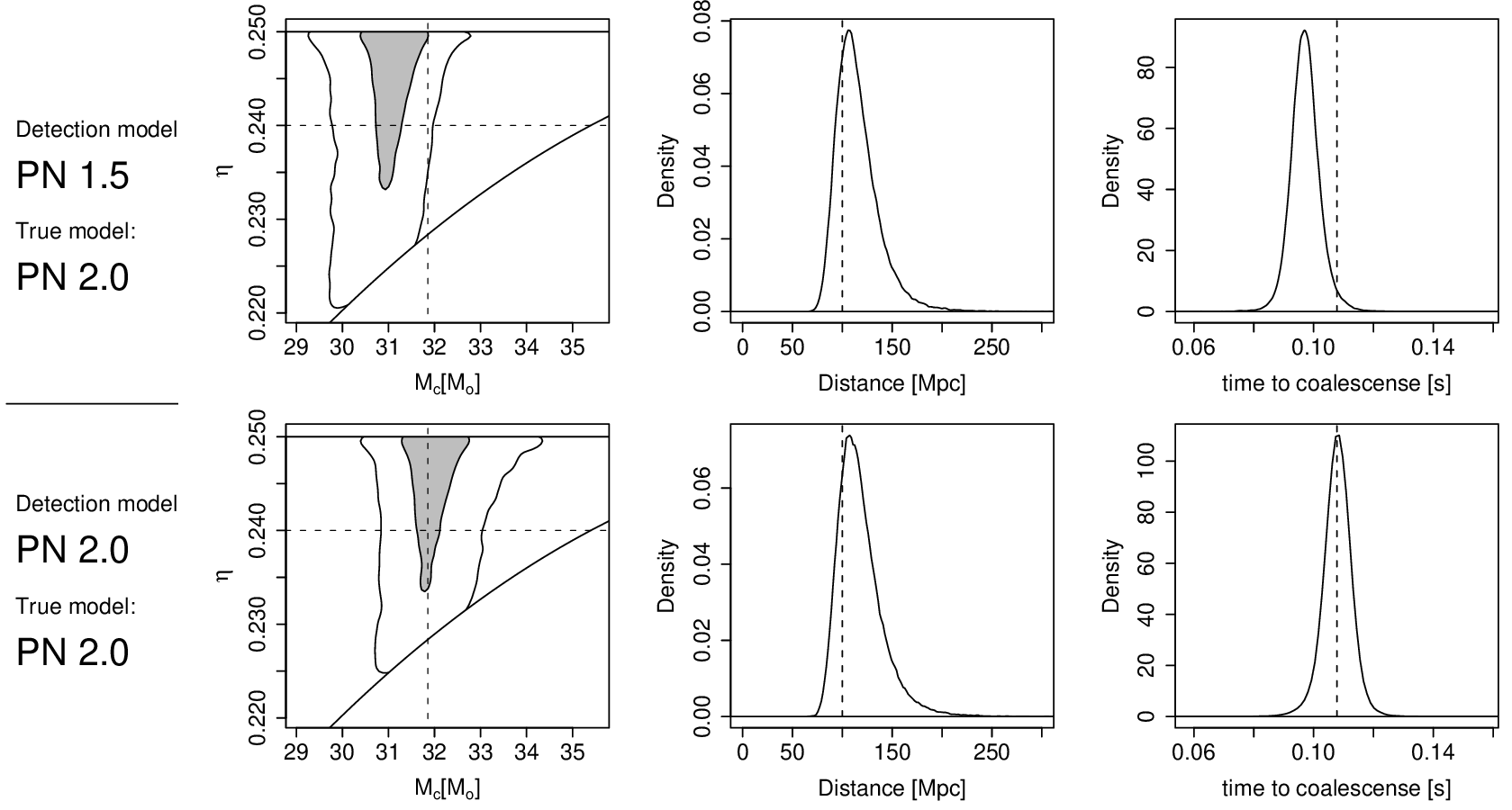}\\
   (a)\\
   \includegraphics[width=7.35cm,angle=270]{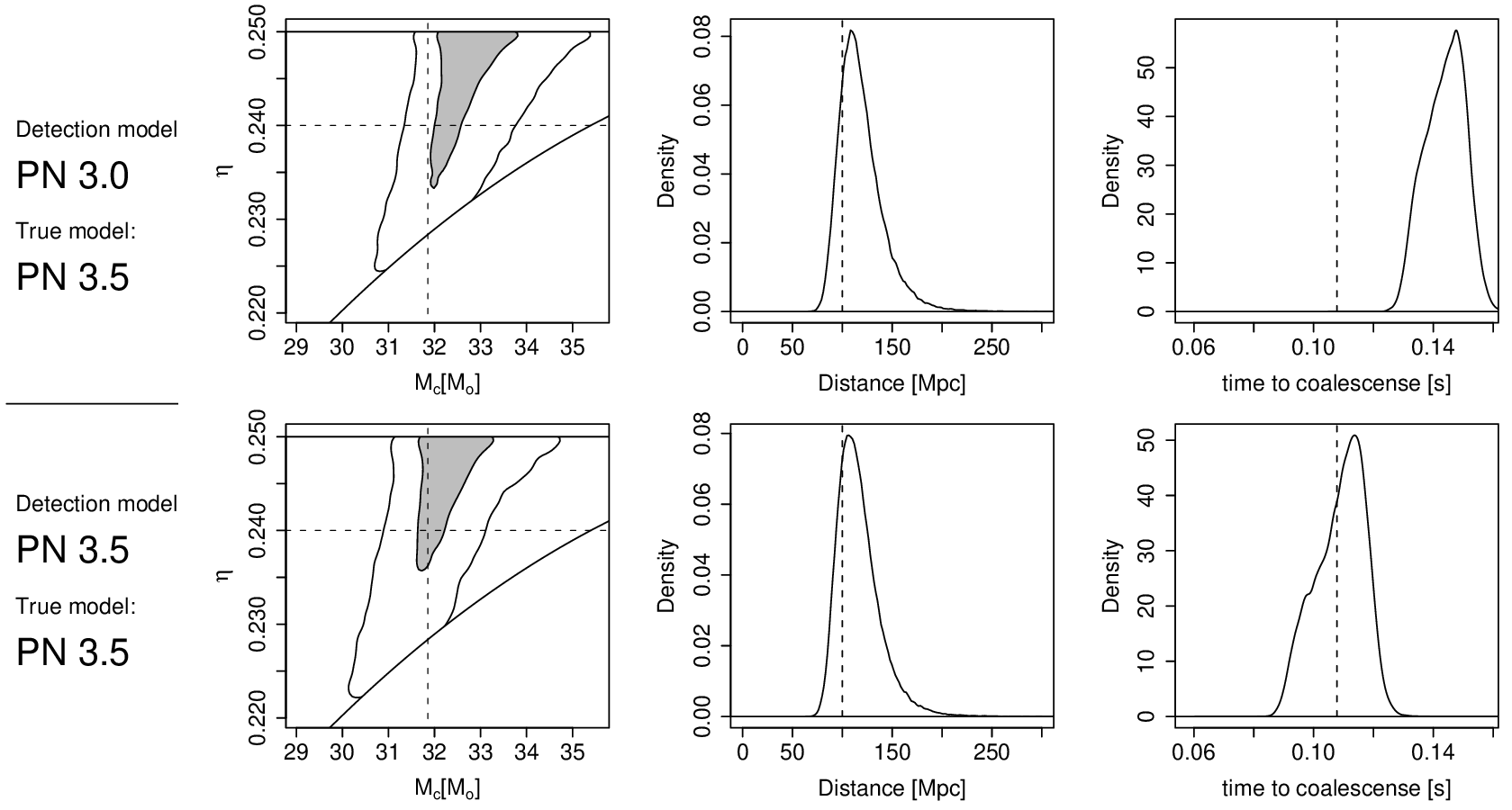} \\
   (b)
  \end{tabular}
\caption{MCMC generated posterior densities for B3. Part (a) shows the comparison with data of a PN2.0 wave form and (b) uses data with the PN3.5 wave form. 
Each of the figures in (a) and (b) show two different model comparisons based either on a lower PN order signal or the same PN order that was used in the data.
The left column shows the joint posterior density of the mass parameters $M_c$ and $\eta$
in form of the 95\% credibility area that contains 95\% of the probability mass and the inner 50\% 
credibility region colored in gray. The middle column shows the MCMC generated kernel density estimate (KDE) of the distance $r$ and the right column the KDE for the time to coalescence $t_C$. The true parameters values are indicated as dashed lines.}
\label{figB3}
  \end{center}
\end{figure}

For the three considered binary systems we generated a total of 36
plots for the distributions of the chirp mass $M_c$ and the mass
function $\eta$, as well as density plots for the distance $r$ and the
time to coalescence $t_C$.  We have chosen to plot the joint posterior
probability of the mass parameters in the $(M_c,\eta)$-space because
they are not as much correlated as $(m_1, m_2)$ in their posterior,
which produce hard to visualize posterior densities.  Although we
could plot the posterior in the $(\lambda_1,\lambda_2)$ space, the
joint posterior probability of $M_c$ and $\eta$ is physically more
meaningful. We divided the 36 plots into three sets corresponding to
B1 (Fig.~\ref{figB1}), B2 (Fig.~\ref{figB2}), and B3 (Fig.~\ref{figB3}).

Figs.~\ref{figB1},~\ref{figB2},~\ref{figB3} display the true
parameter values chosen in our simulations (dashed line in the case of
$r$ and $t_C$, and intersection of two dashed lines in the $M_c, \eta$
plots). From visual inspection we notice that when the model matches
the signal present in the data the posteriors cover well the true
parameter values. In Fig.~\ref{figB1}, however, the joint posterior
distribution of the mass parameters are offset from the true values in
the PN1.5/2.0 comparison. The PN3.0/3.5 detection, on the other
hand, reveals a much smaller offset for this particular signal.
However, this is not true in general, as it can be seen for the
PN3.0/3.5 comparison shown in Fig.~\ref{figB2}. The offsets of the
posterior distributions from the true values of the mass parameters
are very obvious in both, the PN1.5/2.0 and PN3.0/3.5 comparisons.
The posterior is shifted over several of its standard deviation. 
Very striking is also the error in the time
to coalescence for the PN3.0/3.5 comparison in the B2 signal.  The
mass parameters $M_c$ and $\eta$ and time to coalescence are obviously
the parameters subject to biases when using a simplified model. This is
physically understandable since these three parameters define the
phase of the signal. The posterior distributions of the mass-related
parameters shown in Fig.~\ref{figB3} reveal smaller offsets. This is
because the spread of the posterior distribution is less pronounced
and the bias is therefore smaller compared to the posterior standard
deviation. The posterior distribution of time to coalescence, however,
is strongly offset from the true parameter value in the PN3.0/3.5
comparison.

The graphical output only serves as a visualization. For an honest
comparison, numbers are needed.  To this end, in Tab.~\ref{estimatesMc}, 
we show the true values, the 95\%
posterior credibility interval, the median and the mean of the chirp
mass $M_c$, based on the MCMC outputs.  Tab.~\ref{estimateseta},
Tab.~\ref{estimatestc}, and Tab.~\ref{estimatesDist} show the same
entries for the parameters $\eta$, $t_C$, and $r$, respectively. The
right-most column of the tables compares whether the true values of the
binary systems fall into the corresponding 95\% posterior credibility
intervals.

\begin{table}
\begin{tabular}{|c||c|c|c|c|c|}
\hline
simulation&& 95\% credibility  &posterior&posterior& true value falls\\[-3mm]
identification& true value &interval (CI)& mean & median & into 95\% CI\\
\hline
B1: PN1.5/2.0 &$ 0.87905 $&$[ 0.87867 , 0.87978 ]$&$ 0.87908 $&$ 0.87911 $& $\checkmark$ \\ 
B1: PN2.0/2.0 &$ 0.87905 $&$[ 0.87865 , 0.87955 ]$&$ 0.87906 $&$ 0.87907 $& $\checkmark$ \\ 
B1: PN3.0/3.5 &$ 0.87905 $&$[ 0.87855 , 0.87942 ]$&$ 0.87900 $&$ 0.87899 $& $\checkmark$ \\ 
B1: PN3.5/3.5 &$ 0.87905 $&$[ 0.87863 , 0.87942 ]$&$ 0.87905 $&$ 0.87904 $& $\checkmark$ \\ 
\hline
B2: PN1.5/2.0 &$ 3.08951 $&$[ 3.08066 , 3.11047 ]$&$ 3.09384 $&$ 3.09576 $& $\checkmark$ \\ 
B2: PN2.0/2.0 &$ 3.08951 $&$[ 3.07414 , 3.10998 ]$&$ 3.08931 $&$ 3.09047 $& $\checkmark$ \\ 
B2: PN3.0/3.5 &$ 3.08951 $&$[ 3.06665 , 3.09953 ]$&$ 3.07790 $&$ 3.08828 $& $\checkmark$ \\ 
B2: PN3.5/3.5 &$ 3.08951 $&$[ 3.07999 , 3.09852 ]$&$ 3.08955 $&$ 3.08969 $& $\checkmark$ \\ 
\hline
B3: PN1.5/2.0 &$ 31.85576 $&$[ 29.54749 , 32.09530 ]$&$ 31.01141 $&$ 30.97721 $& $\checkmark$ \\ 
B3: PN2.0/2.0 &$ 31.85576 $&$[ 30.76357 , 33.50058 ]$&$ 31.90688 $&$ 31.95160 $& $\checkmark$ \\ 
B3: PN3.0/3.5 &$ 31.85576 $&$[ 31.09861 , 34.68858 ]$&$ 32.53342 $&$ 32.60697 $& $\checkmark$ \\ 
B3: PN3.5/3.5 &$ 31.85576 $&$[ 30.52275 , 33.82286 ]$&$ 32.08128 $&$ 32.07732 $& $\checkmark$ \\ 
\hline
\end{tabular}
\caption{Simulation results of chirp mass $M_c$.}
\label{estimatesMc}
\end{table}
\begin{table}
\begin{tabular}{|c||c|c|c|c|c|}
\hline
simulation&& 95\% credibility  &posterior&posterior& true value falls\\[-3mm]
identification& true value &interval (CI)& mean & median & into 95\% CI\\
\hline
B1: PN1.5/2.0 &$ 0.18750 $&$[ 0.21344 , 0.24412 ]$&$ 0.22521 $&$ 0.22591 $&  \\ 
B1: PN2.0/2.0 &$ 0.18750 $&$[ 0.17844 , 0.20027 ]$&$ 0.18783 $&$ 0.18819 $& $\checkmark$ \\ 
B1: PN3.0/3.5 &$ 0.18750 $&$[ 0.18110 , 0.19046 ]$&$ 0.18559 $&$ 0.18564 $& $\checkmark$ \\ 
B1: PN3.5/3.5 &$ 0.18750 $&$[ 0.18292 , 0.19184 ]$&$ 0.18758 $&$ 0.18753 $& $\checkmark$ \\ 
\hline
B2: PN1.5/2.0 &$ 0.01129 $&$[ 0.01371 , 0.01525 ]$&$ 0.01435 $&$ 0.01449 $&  \\ 
B2: PN2.0/2.0 &$ 0.01129 $&$[ 0.01070 , 0.01210 ]$&$ 0.01129 $&$ 0.01133 $& $\checkmark$ \\ 
B2: PN3.0/3.5 &$ 0.01129 $&$[ 0.01077 , 0.01101 ]$&$ 0.01086 $&$ 0.01095 $&  \\ 
B2: PN3.5/3.5 &$ 0.01129 $&$[ 0.01121 , 0.01137 ]$&$ 0.01129 $&$ 0.01129 $& $\checkmark$ \\ 
\hline
B3: PN1.5/2.0 &$ 0.24000 $&$[ 0.22428 , 0.24998 ]$&$ 0.24350 $&$ 0.24118 $& $\checkmark$ \\ 
B3: PN2.0/2.0 &$ 0.24000 $&$[ 0.22821 , 0.24998 ]$&$ 0.24436 $&$ 0.24245 $& $\checkmark$ \\ 
B3: PN3.0/3.5 &$ 0.24000 $&$[ 0.22796 , 0.24999 ]$&$ 0.24445 $&$ 0.24250 $& $\checkmark$ \\ 
B3: PN3.5/3.5 &$ 0.24000 $&$[ 0.22590 , 0.24999 ]$&$ 0.24426 $&$ 0.24200 $& $\checkmark$ \\ 
\hline
\end{tabular}
\caption{Simulation results of mass ratio $\eta$.}
\label{estimateseta}
\end{table}
\begin{table}
\begin{tabular}{|c||c|c|c|c|c|}
\hline
simulation&& 95\% credibility  &posterior&posterior& true value falls\\[-3mm]
identification& true value &interval (CI)& mean & median & into 95\% CI\\
\hline
B1: PN1.5/2.0 &$ 42.76933 $&$[ 42.76716 , 42.77022 ]$&$ 42.76854 $&$ 42.76856 $& $\checkmark$ \\ 
B1: PN2.0/2.0 &$ 42.76933 $&$[ 42.76805 , 42.77089 ]$&$ 42.76937 $&$ 42.76940 $& $\checkmark$ \\ 
B1: PN3.0/3.5 &$ 42.76933 $&$[ 42.76931 , 42.77313 ]$&$ 42.77113 $&$ 42.77115 $& $\checkmark$ \\ 
B1: PN3.5/3.5 &$ 42.76933 $&$[ 42.76740 , 42.77111 ]$&$ 42.76937 $&$ 42.76935 $& $\checkmark$ \\ 
\hline
B2: PN1.5/2.0 &$ 5.26426 $&$[ 5.23179 , 5.26308 ]$&$ 5.24587 $&$ 5.24686 $&  \\ 
B2: PN2.0/2.0 &$ 5.26426 $&$[ 5.24868 , 5.28231 ]$&$ 5.26398 $&$ 5.26439 $& $\checkmark$ \\ 
B2: PN3.0/3.5 &$ 5.26426 $&$[ 5.52561 , 5.58354 ]$&$ 5.54898 $&$ 5.54958 $&  \\ 
B2: PN3.5/3.5 &$ 5.26426 $&$[ 5.24267 , 5.28606 ]$&$ 5.26438 $&$ 5.26467 $& $\checkmark$ \\ 
\hline
B3: PN1.5/2.0 &$ 0.10778 $&$[ 0.08805 , 0.10761 ]$&$ 0.09706 $&$ 0.09723 $&  \\ 
B3: PN2.0/2.0 &$ 0.10778 $&$[ 0.09796 , 0.11714 ]$&$ 0.10802 $&$ 0.10794 $& $\checkmark$ \\ 
B3: PN3.0/3.5 &$ 0.10778 $&$[ 0.13103 , 0.15565 ]$&$ 0.14444 $&$ 0.14390 $&  \\ 
B3: PN3.5/3.5 &$ 0.10778 $&$[ 0.09220 , 0.12211 ]$&$ 0.10999 $&$ 0.10884 $& $\checkmark$ \\ 
\hline
\end{tabular}
\caption{Simulation results of time to coalescence $t_C$.}
\label{estimatestc}
\end{table}
\begin{table}
\begin{tabular}{|c||c|c|c|c|c|}
\hline
simulation&& 95\% credibility  &posterior&posterior& true value falls\\[-3mm]
identification& true value &interval (CI)& mean & median & into 95\% CI\\
\hline
B1: PN1.5/2.0 &$ 16.00 $&$[ 14.04 , 21.63 ]$&$ 17.00 $&$ 17.22 $& $\checkmark$ \\ 
B1: PN2.0/2.0 &$ 16.00 $&$[ 14.03 , 21.55 ]$&$ 16.96 $&$ 17.17 $& $\checkmark$ \\ 
B1: PN3.0/3.5 &$ 16.00 $&$[ 14.18 , 22.04 ]$&$ 17.20 $&$ 17.44 $& $\checkmark$ \\ 
B1: PN3.5/3.5 &$ 16.00 $&$[ 14.15 , 21.92 ]$&$ 17.18 $&$ 17.40 $& $\checkmark$ \\ 
\hline
B2: PN1.5/2.0 &$ 22.00 $&$[ 19.13 , 31.17 ]$&$ 23.67 $&$ 24.07 $& $\checkmark$ \\ 
B2: PN2.0/2.0 &$ 22.00 $&$[ 19.02 , 30.98 ]$&$ 23.61 $&$ 23.98 $& $\checkmark$ \\ 
B2: PN3.0/3.5 &$ 22.00 $&$[ 19.23 , 31.83 ]$&$ 23.86 $&$ 24.59 $& $\checkmark$ \\ 
B2: PN3.5/3.5 &$ 22.00 $&$[ 19.23 , 31.35 ]$&$ 23.71 $&$ 24.11 $& $\checkmark$ \\ 
\hline
B3: PN1.5/2.0 &$ 100.00 $&$[ 84.54 , 168.42 ]$&$ 111.55 $&$ 115.45 $& $\checkmark$ \\ 
B3: PN2.0/2.0 &$ 100.00 $&$[ 86.07 , 174.08 ]$&$ 114.05 $&$ 118.16 $& $\checkmark$ \\ 
B3: PN3.0/3.5 &$ 100.00 $&$[ 86.94 , 172.77 ]$&$ 114.94 $&$ 118.91 $& $\checkmark$ \\ 
B3: PN3.5/3.5 &$ 100.00 $&$[ 85.92 , 169.58 ]$&$ 113.05 $&$ 116.76 $& $\checkmark$ \\ 
\hline
\end{tabular}
\caption{Simulation results of distance $r$.}
\label{estimatesDist}
\end{table}
The results seen in these tables show that the mass function, $\eta$,
and the time to coalescence, $t_C$, are the parameters that are most
biased when estimated with a simplified model.  Note, first of all,
that in all cases, the true values fall into the 95\% credibility
intervals when the estimation is based on the true model.  For the
distance $r$ and the chirp mass $M_c$ the 95\% credibility intervals
cover the true values in all comparisons.  However, when applying a
simplified model, the 95\% credibility interval of the mass function
$\eta$ does not cover the true value in three cases: \{B1:PN1.5/2.0,
B2:PN1.5/2.0, and B2:PN3.0/3.5)\}. In the case of the time to
coalescence $t_C$, we find instead four cases:
\{B2:PN1.5/2.0,B2:PN3.0/3.5,B3:PN1.5/2.0,B3:PN3.0/3.5\}.  Combining
these cases, we have a total of 5 out of the 6 simple model
comparisons (PN1.5/2.0 and PN3.0/3.5) that fail to retrieve all their
parameters within the 95\% credibility region. The only simple model
comparison that yields 95\% credibility intervals that overlap all the
true parameter values is B1: PN1.5/2.0, although this is only
marginal.

In summary, we see that the bias in the estimated parameters based on
a simpler model is larger than the statistical uncertainty.  However,
we should note that the SNR we have been considering corresponds to
the value at which the posterior detection probability just reaches
the value of one. Since the statistical error is a monotonically
decreasing function of the SNR while the bias is not, we conclude that
the difference between statistical and systematic error increases for
larger SNRs. 

These results reveal that parameter estimates based on simplified
models are not very reliable, since the systematic error is higher
than the uncertainty of the posterior distribution. Furthermore, the
use of higher order post Newtonian wave forms does not abate this
problem, as it has been shown in Fig.~\ref{figB2}.

\section{Conclusion}
\label{SecVI}
We have shown that, within the Bayesian framework, the probability of
detection is not impeded by using a simplified model for detecting
wave forms of higher PN order in the low-SNR regime. The Bayesian
approach provides the means to gain insight into the variation of the
detection probability over different noise realizations. We have shown
that the difference between the posterior detection probabilities
corresponding to the true and the simplified model is very small as
compared to its variance over different noise realizations.  We
have further shown that the systematic error in the Bayesian
estimates, on the other hand, can be larger than the statistical
uncertainties.  
This is also in agreement with results obtained within
the frequentist approach, discussed in the literature by others 
\cite{Chronopoulos:2001,Canitrot:2001,PBBCKWPM07}.
However, it is based on finding the best fit of the involved wave forms while
in our Bayesian framework an integration is performed over the entire 
posterior distribution which implies detection and estimation simultaneously.
We can therefore analyze the posterior distributions, conditioned
on the model that involves a signal which provides us with credible estimates
in the low-SNR regime. 

We find that the estimates of $\eta$ and $t_C$ based on simplified models need to be taken with caution
in both the PN1.5/2.0 and in the PN3.0/3.5 case as the offset is unpredictable.
The only parameter that is accurately recovered throughout our simulations is the
distance $r$ which is clear as it only appears in the amplitude term and is not affecting the phase evolution of
the signal. 
The chirp mass could also be retrieved within the 95\% credibility intervals
but yet shows a visible offset. With increasing SNR, however, the statistical error becomes smaller while
the systematical offset remains constant.
Given these findings we conclude that post Newtonian approximations, regardless of order,
can be precarious for detecting ``true'' gravitational wave forms. 

\section{Acknowledgments}

This research was performed at the Jet Propulsion Laboratory,
California Institute of Technology, under contract with the National
Aeronautics and Space Administration.  The supercomputers used in this
investigation were provided by funding from JPL Office of the Chief
Information Officer.  This work was supported by the NASA Postdoctoral
program Fellowship appointment conducted at JPL (R.U.), and by the
research task 05-BEFS05-0014 (M.T.).

\bibliographystyle{unsrt}

\begin{thebibliography}{10}

\bibitem{LIGO}
The {LIGO} project home page is at http://www.ligo.caltech.edu.

\bibitem{VIRGO}
The virgo project home page is at http://www.virgo.infn.it/.

\bibitem{GEO}
The geo-600 project home page is at http://www.geo600.uni-hannover.de/.

\bibitem{TAMA}
The tama project home page is at http://tamago.mtk.nao.ac.jp/.

\bibitem{Helstrom68}
C.W. Helstr\"om, editor.
\newblock {\em Statistical Theory of Signal Detection}.
\newblock Pergamon Press, London, 1968.

\bibitem{Thorne87}
K.S. Thorne.
\newblock Gravitational radiation.
\newblock In {\em Three Hundred Years of Gravitation}, pages 330--458.
  Cambridge University Press, 1987.
\newblock S.W. Hawking, and W. Israele, eds.

\bibitem{P05}
F.~Pretorius.
\newblock Evolution of binary black-hole spacetimes.
\newblock {\em Physical Review Letters}, 95:121101, 2005.

\bibitem{CLMZ06}
M.~Campanelli, C.O. Lousto, P.~Marronetti, and Y.~Zlochower.
\newblock Accurate evolutions of orbiting black-hole binaries without excision.
\newblock {\em Physical Review Letters}, 96:111102, 2006.

\bibitem{BCCKM06}
J.G. Baker, J.~Centrella, D.~Choi, M.~Koppitz, and J.~van Meter.
\newblock Gravitational-wave extraction from an inspiraling configuration of
  merging black holes.
\newblock {\em Physical Review Letters}, 96:111102, 2006.

\bibitem{BCP06}
A.~Buonanno, G.~B. Cook, and F.~Pretorius.
\newblock Inspiral, merger, and ring-down of equal-mass black-hole binaries.
\newblock {\em Physical Review D}, 75:124018, 2006.

\bibitem{BMWCK06}
J.G. Baker, J.R. van Meter, S.T. McWilliams, J.~Centrella, and B.J. Kelly.
\newblock Consistency of post-newtonian waveforms with numerical relativity.
\newblock {\em Classical and Quantum Gravity}, pages arXiv:gr--qc/0612024v1,
  2006.

\bibitem{PBBCKWPM07}
Y.~Pan, A.~Buonanno, J.G. Baker, J.~Centrella, B.J. Kelly, S.T. McWilliams,
  F.~Pretorius, and J.R. van Meter.
\newblock A data-analysis driven comparison of analytic and numerical
  coalescing binary waveforms: nonspinning case.
\newblock {\em Classical and Quantum Gravity}, page arXiv:0704.1964v1, 2006.

\bibitem{CutVallis:2007}
C.~Cutler and M.~Vallisneri.
\newblock Lisa detections of massive black hole inspirals: parameter extraction
  errors due to inaccurate template waveforms.
\newblock {\em Classical and Quantum Gravity}, page arXiv:0707.2982, 2007.

\bibitem{Vallis:2007}
M.~Vallisneri.
\newblock Use and abuse of the fisher information matrix in the assessment of
  gravitational-wave parameter-estimation prospects.
\newblock {\em Classical and Quantum Gravity}, pages arXiv:gr--qc/0703086,
  2007.

\bibitem{Gilks:1996}
W.~R. Gilks, S.~Richardson, and P.~J. Spiegelhalter, D. J.~Green, editors.
\newblock {\em {M}arkov chain {M}onte {C}arlo in practice}.
\newblock Chapman and Hall, London, 1996.

\bibitem{RoeverMeyerChristensen:2006}
C.~R\"{o}ver, R.~Meyer, and N.~Christensen.
\newblock Bayesian inference on compact binary inspiral gravitational radiation
  signals in interferometric data.
\newblock {\em Classical and Quantum Gravity}, 23(15):4895--4906, August 2006.

\bibitem{RoeverMeyerChristensen:2007}
C.~R\"{o}ver, R.~Meyer, and N.~Christensen.
\newblock Coherent {B}ayesian inference on compact binary inspirals using a
  network of interferometric gravitational wave detectors.
\newblock {\em Physical Review~D}, 75(6):062004, March 2007.

\bibitem{RoeverMeyerGuidiVicereChristensen:2007}
C.~R\"{o}ver, R.~Meyer, G.~M. Guidi, A.~Vicer\'e, and N.~Christensen.
\newblock Coherent {B}ayesian analysis of inspiral signals.
\newblock {\em Classical and Quantum Gravity}, 24(19):S607--S615, October 2007.

\bibitem{Jaynes:2003}
E.~T. Jaynes.
\newblock {\em Probability Theory: The Logic of Science}.
\newblock Cambridge University Press, 2003.

\bibitem{Loredo:1992}
T.~J. Loredo.
\newblock {\em Statistical challanges in modern astronomy}, chapter The promise
  of {B}ayesian inference for astrophysics, pages 275--297.
\newblock Springer-Verlag, New York, 1992.

\bibitem{Gregory:2005}
P.~C. Gregory, editor.
\newblock {\em Bayesian Logical Data Analysis for the Physical Sciences}.
\newblock Cambridge University Press, 2005.

\bibitem{Sandage:2001}
A.~Sandage.
\newblock {\em Encyclopedia of Astronomy and Astrophysics}.
\newblock Institute of Physics Publishing, Bristol, 2001.

\bibitem{PDB01}
A.~Pai, S.~Dhurandhar, and S.~Bose.
\newblock Data-analysis strategy for detecting gravitational-wave signals from
  inspiraling compact binaries with a network of laser-interferometric
  detectors.
\newblock {\em Physical Review D}, 64:042004, 2001.

\bibitem{Kass:1995}
R.~E. Kass and A.~E. Raftery.
\newblock Bayes factors.
\newblock {\em Journal of the American Statistical Association},
  90(430):773--795, 1995.

\bibitem{Han:2001}
C.~Han and B.~P. Carlin.
\newblock {M}arkov chain {M}onte {C}arlo methods for computing Bayes factors: A
  comparative review.
\newblock {\em Journal of the American Statistical Association},
  96(455):1122--1132, 2001.

\bibitem{Schwarz:1978}
G.~Schwarz.
\newblock Estimating the dimension of a model.
\newblock {\em The Annals of Statistics}, 6(2):461--464, 1978.

\bibitem{Green:1995}
P.~J. Green.
\newblock Reversible jump {M}arkov chain {M}onte {C}arlo computation and
  {B}ayesian model determination.
\newblock {\em Biometrika}, 82(4):711--732, 1995.

\bibitem{GreenHjortRichardson:2003}
P.~J. Green, N.~L. Hjort, and S.~Richardson, editors.
\newblock {\em Highly Structured Stochastic Systems}.
\newblock Oxford University Press, 2003.

\bibitem{Marinari:1992}
E.~Marinari and G.~Parisi.
\newblock Simulated tempering: A new {M}onte {C}arlo scheme.
\newblock {\em Europhysics Letters}, 19:451--454, 1992.

\bibitem{Neal:1996}
R.~M. Neal.
\newblock Sampling from multimodal distributions using tempered transitions.
\newblock {\em Statistics and Computing}, 6:353--366, 1996.

\bibitem{Geyer:1991}
C.~J. Geyer.
\newblock {M}arkov chain {M}onte {C}arlo maximum likelihood.
\newblock In E.~M. Keramidas, editor, {\em Computing science and statistics:
  Proceedings of the 23rd Symposium on the Interface}, pages 156--163.
  Interface Foundation, Fairfax Station, VA, 1991.

\bibitem{UmstaetterThesis:2006}
R.~Umstaetter.
\newblock {\em Bayesian Strategies for Gravitational Radiation Data Analysis}.
\newblock PhD thesis, University of Auckland, 2006.

\bibitem{Hansmann:1997}
U.~H.~E. Hansmann.
\newblock Parallel tempering algorithm for conformational studies of biological
  molecules.
\newblock {\em Chemical Physics Letters}, 281:140--150, 1997.

\bibitem{Metropolis:1953}
N.~Metropolis, A.~W. Rosenbluth, M.~N. Rosenbluth, A.~H. Teller, and E.~Teller.
\newblock Equation of state calculations by fast computing machines.
\newblock {\em Journal of Chemical Physics}, 21(6):1087--1092, 1953.

\bibitem{Hastings:1970}
W.~K. Hastings.
\newblock {M}onte {C}arlo sampling methods using {M}arkov chains and their
  applications.
\newblock {\em Biometrika}, 57:97--109, 1970.

\bibitem{UMDVWC:2004}
R.~Umst\"atter, R.~Meyer, R.J. Dupuis, J.~Veitch, G.~Woan, and N.~Christensen.
\newblock Estimating the parameters of gravitational waves from neutron stars
  using an adaptive {MCMC} method.
\newblock {\em Classical and Quantum Gravity}, 21:S1655--S1665, 2004.

\bibitem{MaxEnt:2004}
R.~Umst\"atter, R.~Meyer, R.J. Dupuis, J.~Veitch, G.~Woan, and N.~Christensen.
\newblock Detecting gravitational radation from neutron stars using a
  six-parameter adaptive {MCMC} method.
\newblock In {\em AIP Conference Proceedings - {B}ayesian inference and Maximum
  Entropy Methods in Science and Engineering: 24th}, volume 735, pages
  336--343. American Institute of Physics, 2004.
\newblock International Workshop on {B}ayesian Inference and Maximum Entropy
  Methods in Science and Engineering.

\bibitem{Chronopoulos:2001}
A.E. Chronopoulos and T.A. Apostolatos.
\newblock Less accurate but more efficient family of search templates for
  detection of gravitational waves from inspiraling compact binaries.
\newblock {\em Physical Review D}, 64:042003, 2001.

\bibitem{Arun:2005}
K.~G. Arun, B.~R. Iyer, B.~S. Sathyaprakash, and P.~A. Sundararajan.
\newblock Parameter estimation of inspiralling compact binaries using 3.5
  post-newtonian gravitational wave phasing: The nonspinning case.
\newblock {\em Physical Review D}, 71:084008, 2005.

\bibitem{Canitrot:2001}
P.~Canitrot.
\newblock Systematic errors for matched filtering of gravitational waves from
  inspiraling compact binaries.
\newblock {\em Physical Review D}, page 082005, 2001.

\end{thebibliography}

\end{document}